\documentclass[compsoc, conference, a4paper, 10pt, times]{IEEEtran}
\IEEEoverridecommandlockouts
\usepackage{cite}
\usepackage{amsmath,amssymb,amsfonts}

\usepackage{microtype}
\usepackage{balance}
\usepackage{hyperref}

\usepackage{algorithmic}
\usepackage{graphicx}
\usepackage{textcomp}
\usepackage{makecell}
\usepackage{xcolor}
\def\BibTeX{{\rm B\kern-.05em{\sc i\kern-.025em b}\kern-.08em
    T\kern-.1667em\lower.7ex\hbox{E}\kern-.125emX}}

\usepackage[disable]{todonotes}
\usepackage{multirow}
\usepackage{booktabs}

\usepackage[absolute,showboxes]{textpos}

\usepackage{caption}
\usepackage[labelformat=simple]{subcaption}

\usepackage{url} %

\usepackage{arydshln}

\newcommand{\result}[1]{}

\definecolor{myred}{cmyk}{0, 0.7808, 0.4429, 0.1412}

\newcommand{\done}[1]{}

\usepackage{pifont}
\newcommand{\cmark}{\ding{51}}%
\newcommand{\xmark}{\ding{56}}%

\usepackage{xspace}

\newcommand{\eg}{\textit{e.g.,}~}
\newcommand{\ie}{\textit{i.e.,}~}
\newcommand{\etc}{\textit{etc.}\xspace}
\newcommand{\one}{({\em i})\xspace}
\newcommand{\two}{({\em ii})\xspace}
\newcommand{\three}{({\em iii})\xspace}
\newcommand{\four}{({\em iv})\xspace}

\newcommand*\circled[1]{\tikz[baseline=(char.base)]{
            \node[shape=circle,draw,inner sep=0.9pt] (char) {\sf \small #1};}}

\makeatletter
\renewcommand{\paragraph}[1]{\vspace*{0.03in}\noindent{\bf #1.}\hspace{0.25ex \@plus1ex \@minus.2ex}}
\newcommand{\paragraphNoDot}[1]{\vspace*{0.03in}\noindent{\bf #1}\hspace{0.25ex \@plus1ex \@minus.2ex}}
\makeatother

\newcommand*\dhline{\specialrule{0pt}{1pt}{0pt}\hdashline[.4pt/3pt]\specialrule{0pt}{0pt}{2pt}}

\begin{document}

\definecolor{boxgray}{rgb}{0.93,0.93,0.93}
\textblockcolor{boxgray}
\setlength{\TPboxrulesize}{0.4pt}
\setlength{\TPHorizModule}{\paperwidth}
\setlength{\TPVertModule}{\paperheight}
\TPMargin{5pt}
\begin{textblock}{0.8}(0.1,0.02)
	\noindent
	\footnotesize
	\centering
	If you cite this paper, please use the Euro S\&P reference:\\
	M. Nawrocki, J. Kristoff, R. Hiesgen, C. Kanich,  T. C. Schmidt, and M. Wählisch.
	2023.\\ SoK: A Data-driven View on Methods to Detect Reflective Amplification DDoS Attacks Using Honeypots.\\
	\emph{In Proceedings of Euro S\&P ’23.} IEEE, Piscataway, NJ, USA.
\end{textblock}

\title{SoK: A Data-driven View on Methods to \\ Detect Reflective Amplification DDoS Attacks Using Honeypots}

\author{
\IEEEauthorblockN{Marcin Nawrocki\IEEEauthorrefmark{1},
  John Kristoff\IEEEauthorrefmark{3}\IEEEauthorrefmark{4},
  Raphael Hiesgen\IEEEauthorrefmark{5},
  Chris Kanich\IEEEauthorrefmark{3},
  Thomas C. Schmidt\IEEEauthorrefmark{5},
  Matthias W\"ahlisch\IEEEauthorrefmark{2}\IEEEauthorrefmark{1}
}
\IEEEauthorblockA{
  \IEEEauthorrefmark{1}Freie Universit\"at Berlin,
  \IEEEauthorrefmark{2}TU Dresden,
  \IEEEauthorrefmark{3}University of Illinois at Chicago,
  \IEEEauthorrefmark{4}NETSCOUT,
  \IEEEauthorrefmark{5}HAW Hamburg
}
\IEEEauthorblockN{
  \texttt{marcin.nawrocki@fu-berlin.de},
  \texttt{\{jkrist3, ckanich\}@uic.edu}, 
  \\
  \texttt{\{raphael.hiesgen, t.schmidt\}@haw-hamburg.de},
  \texttt{matthias.waehlisch@tu-dresden.de}
}
}

\maketitle

\begin{abstract}

In this paper, we revisit the use of honeypots for detecting reflective amplification attacks.
These measurement tools require careful design of both data collection and data analysis including cautious threshold inference.
We survey common amplification honeypot platforms as well as the underlying methods to infer attack detection thresholds and to extract knowledge from the data.
By systematically exploring the threshold space, we find most honeypot platforms produce comparable results despite their different configurations.
Moreover, by applying data from a large-scale honeypot deployment, network telescopes, and a real-world baseline obtained from a leading DDoS mitigation provider, we question the fundamental assumption of honeypot research that convergence of observations can imply their completeness.
Conclusively we derive guidance on precise, reproducible honeypot research, and present open challenges.

\end{abstract}


\section{Introduction}
\label{sec:intro}

Distributed Denial of Service (DDoS) attacks are a serious threat to the
Internet infrastructure.  Reflective amplification
attacks~\cite{r-ahrnp-14,rowrs-adads-15}, a specific DDoS type, are a unique burden
since they allow an attacker to trigger large traffic volumes from third
parties by exploiting protocol mechanics rather than hijacking hosts.
Over the last many years, amplification attacks have been responsible
for a significant number of attacks~\cite{netscout2019report}.

A common approach to detect amplification attacks in the wild is
the deployment of honeypots~\cite{nawrocki2016survey}. They mimic application protocols such as DNS and NTP that are susceptible to amplification attacks, wait for
attackers to interact, and then log attack traffic attempting to abuse
them as amplifiers.  Amplification honeypots may be able to infer the size of attacks
based on traffic patterns as well as identify the victims they are
instructed to reflect toward.

Research on amplification honeypots is usually guided by three questions to
evaluate whether honeypots are a viable tool. First, which heuristics
identify packets that correspond to an attack in a train of
packets captured by honeypots (\emph{attack detection})? Second, how many
honeypot sensors are necessary to capture a stable amount of events
(\emph{honeypot convergence})? Third, do sensors capture a representative
view of Internet-wide attacks (\emph{completeness})? These aspects should
be considered separately. Attack detection, for example, might be accurate
on a given data set, while the data set does not include all attacks.

Surprisingly, our community mixes detection, convergence, and completeness.
For more than ten years, we have been holding the common belief ``[t]he more honeypots we deploy, the more likely one of them is contacted''~\cite{ph-vhfbt-08}.
Even with the advent of amplification honeypots we still believe that we can nearly achieve completeness: ``This shows that---per mode---we had enough honeypots to cover most attacks out there.''\cite{kramer2015amppot}, ``[\dots] as many as 150 honeypots are needed to capture 99\% of actor behavior''~\cite{griffioen2021adversarial}, ``[\dots] our reflectors can see between 85.1\% and 96.6\% of UDP reflection attacks''~\cite{thomas2017thousand}.
A key insight of this paper is that reality is different.

\begin{figure}
  \center
  \includegraphics[trim={2cm 2cm 2cm 3cm},clip,width=0.9\columnwidth]{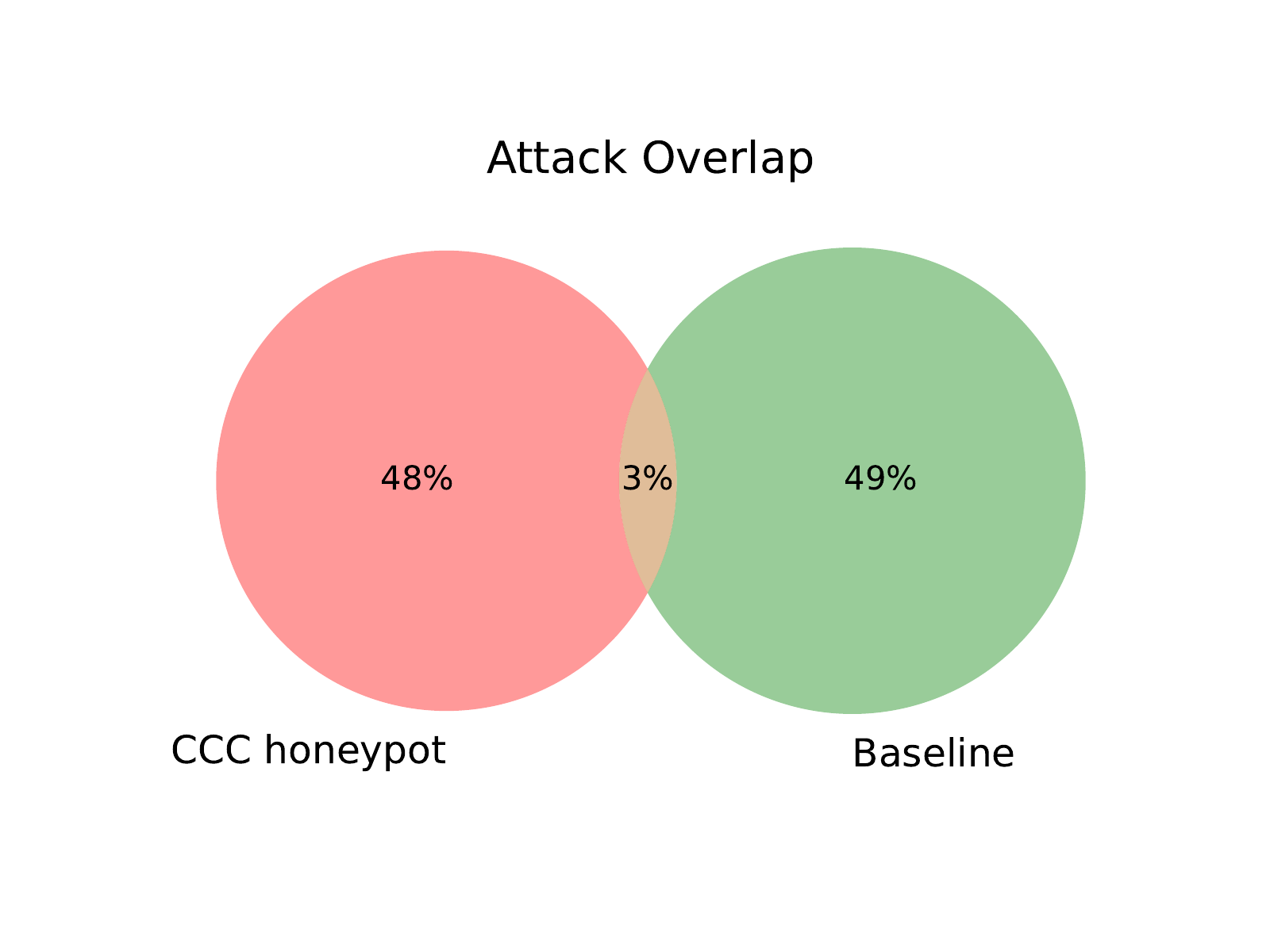}
	\caption{Relative shares of victims observed at a large-scale amplification honeypot and confirmed at a large DDoS mitigation provider (Baseline).}
  \label{fig:overlap-venn}
\end{figure}

In this paper, we revisit the long-held beliefs about the visibility and attack detection precision of honeypots.
We combine two different methods by \one systematizing and contextualizing existing knowledge and \two implementing a data-driven approach, which clearly shows that common beliefs do not hold.

Based on an extensive literature study, we select six amplification honeypots and compare them.
The six honeypot platforms were used in security studies when analyzing reflective-amplification attacks based on honeypot data.
They have been published, cited recently, and had a notable impact on security research.
We implement three steps.
\one We survey the honeypot deployment configurations that enable observations, \eg the number of honeypot sensors deployed and the geographical and topological distribution of the platform,
\two we describe the attack definitions that are used to understand the observations, %
and
\three we assess the rationale behind the argument that the deployed honeypot achieves completeness. %

To bolster our arguments, we conduct a data-driven approach.
Our data corpus covers three months and includes measurements from a large-scale honeypot, four network telescopes distributed in the US and Europe, and baseline real-world alerts from a leading DDoS mitigation provider.
\autoref{fig:overlap-venn} motivates this approach. It shows the overlap
of victims under attack monitored by a well-known research honeypot project and
a baseline of attacks against customers of a leading DDoS mitigation provider.
The overlap is small, and most importantly the honeypots do not capture a
significant portion of attacks targeting real-world networks, even though a
honeypot could capture those incidents in principle.

\paragraph{Contributions}
In a nutshell, our systematization of knowledge stresses that the research community could benefit from a framework that allows for algorithmic assessment of honeypot deployments and, to assemble packets captured by honeypots to malicious flows, from attack detection heuristics that adaptively incorporate deployment properties.
Our key contributions are:
\begin{enumerate}
 \item We explore the comparability of the attack detection thresholds used by six honeypot
    platforms, and place them in the complete threshold space. All thresholds but one produce
    similar results.

 \item We present a systematic approach to analyze data collected by honeypots. We
    identify the key properties that should be considered and documented to improve reproducibility of future honeypot research.

 \item We show that honeypot convergence, a frequently used measure, is a poor
 indicator for the completeness of observations. This metric is statistically
 unstable. Sizable honeypot platforms only observe up to 11\% of baseline
 attacks.

 \item We find that current honeypot deployments do not significantly benefit from better
    attack detection thresholds because attackers simply do not interact with
    honeypots. This may help to improve the placement of honeypot sensors in the future.

 \item We discuss how amplification features of protocols can influence honeypot
    observations and analysis.

\end{enumerate}

\paragraph{Outline}
The remainder of this paper is guided by our research questions, see \autoref{tab:paper_structure}.
We present basic background in \autoref{sec:background}, introduce our method in \autoref{sec:methodology}, and survey common honeypot platforms in \autoref{sec:honeypot:projects}.
In \autoref{sec:data-sets}, we present the data sets that we use for our data-driven analysis,
We revisit attack detection, convergence, and completeness in \autoref{sec:thresholds} to \autoref{sec:completeness}.
In \autoref{sec:ecosystem}, we present further deployment dimensions of
honeypots. %
We discuss our findings comprehensively and provide further guidance in \autoref{sec:discussion}, and conclude in \autoref{sec:conclusion}.

\begin{table}%
\setlength{\tabcolsep}{2.5pt}
  \caption{Our SoK addresses the following research questions, guiding \one knowledge contextualization, \two data-driven evaluation, and \three further discussions.}
  \label{tab:paper_structure}
  \centering
  \begin{tabular}{l l r}
    \toprule
      SoK & Research Question & Section \\
    \midrule
        Introduce & Which kinds of attacks and monitoring exist? & \ref{sec:background} \\
    \midrule
        Compare & How are amplification honeypots deployed? & \ref{sec:honeypot:projects} \\
        Compare & How are attacks inferred? & \ref{sec:thresholds},\ref{sec:thresholds:current_methods} \\
        Compare & How are comprehensive measurements justified? & \ref{sec:ecosystem:protocols},\ref{sec:convergence} ,\ref{sec:convergence:current_methods} \\
    \midrule
        Evaluate & Do different attack thresholds skew the results? & \ref{sec:thresholds:comparability} \\
        Evaluate & Do honeypots observe all attacks? & \ref{sec:convergence:reproducing},\ref{sec:convergence:fair_introspection},\ref{sec:convergence:vs_completeness},\ref{sec:precision:completeness} \\
        Evaluate & Do we need more precise attack thresholds? & \ref{sec:precision:better_atk_thresholds},\ref{sec:precision:scan_thresholds} \\
    \midrule
        Discuss & What makes measurements prone to errors? & \ref{sec:ecosystem} \\
        Discuss & What do we recommend for future work? & \ref{sec:discussion} \\
	\bottomrule
  \end{tabular}
\end{table}
\setlength{\tabcolsep}{5.25pt}

\section{Problem Statement and Background}
\label{sec:background}

\subsection{Distributed Denial of Service (DDoS) Attacks}
\label{sec:background:attacks}

Denial of Service (DoS) attacks impair the network availability of their
victims.  This is achieved by resource exhaustion caused by overloading
the infrastructure with excessive traffic volume or connection state at
the victim.  Attackers either set up genuine communication channels with
the victims or spoof IP source addresses to obfuscate their attacks.
Both methods are typically conducted using a distributed botnet.  Two
attack types exist, each of which take advantage of the first round-trip
time when a server responds to client requests.

\paragraphNoDot{\one State-building, \textbf{randomly-spoofed attacks}} such as TCP SYN or QUIC Initial floods.  Each spoofed request
can trick the server into setting up a new connection context for
non-existent clients. The network stack will
maintain all currently active connections, including those from spoofed
sources, which fill up the connection queues and cause legitimate
requests to fail.  Since the server tries to respond to each connection
request, it will send \textit{backscatter}, \eg TCP SYN/ACK or QUIC
(server-) Initial packets, to the spoofed addresses. TCP SYN cookies and
QUIC RETRYs may mitigate those attacks~\cite{syncookies,nawrocki2021quic}.

\paragraphNoDot{\two \textbf{Distributed Reflective amplification
attacks}} (DRDoS) combine targeted address spoofing and protocol
mechanics of public services such as DNS and NTP to amplify response
traffic to the victim.  In a DRDoS attack, request packets with the
spoofed source address of the victim are sent to public third-party
servers.  These servers act as amplifiers since responses to the victim
can be be many times larger than the original request~\cite{r-ahrnp-14}.
For example, a typical DNS query packet is about 100 bytes, but a
response to an \texttt{IN ANY} query can often exceed 2000 bytes in
practice.  Attackers seek to minimize the request volume towards
amplifiers whilst maximizing the response volume reflected to the
victim.  This may congest network links along the path to the victim. 

\paragraph{Attack Popularity over Time} Conceptually, DRDoS was already utilized in 1997 with ICMP smurf attacks.
However, direct-path SYN-floods remained the most popular DDoS attack vector from 1996 to 2018 and were then overtaken by DNS reflection-amplification in 2018. %
This popularity was due to 
\one the commercialization of this attack type by booter services, making it available to the non-tech-savvy public, and
\two easier and faster detection of amplifiers based on ready-to-use tools implementing state-less, Internet-wide scans.

\subsection{Honeypots and Network Telescopes}
\label{sec:background:honeypots}

\paragraph{Honeypots} Honeypots are decoy computer resources whose value
lies in being probed, inciting interaction with attackers, and possibly
getting compromised~\cite{ph-vhfbt-08}. They are not a preventive
countermeasure such as firewalls but a way to  detect the
presence of actions that harm a system.  Since honeypots do not offer
production-critical services, all connections to the honeypot are
inherently suspicious.  This enables easy detection of an unauthorized
probe, scan, or attack, because malicious actions are not buried in the
vast amount of legitimate production activities.

Honeypots can be classified along two dimensions, based on the level and
type of interaction they offer.  First, based on the level of
interaction the delineation is \one low-interaction honeypots, \two
medium-interaction honeypots and \three high-interaction honeypots.
Low-interaction honeypots offer only a minimal response-behavior, \eg
they only perform transport-layer handshakes.  Medium-interaction
honeypots extend this behavior by emulating vulnerable services or
partially exposing vulnerable components, \ie they produce valid replies
for specific applications.  Given the reduced interaction capabilities
in low- and medium-interaction honeypots, the chances of compromise are
minimal, which eases deployment.  High-interaction honeypots offer
unrestricted, real operating system environments.  They are more complex
to implement, deploy, and maintain.  They enable, however, forensics to
fully observe the behavior of malware, \eg bots, or ransomware.

Second, based on the type of interaction they offer, honeypots are
classified into \one server and \two client honeypots.  Server honeypots
wait for an incoming connection.  They may not advertise services
explicitly, more likely they are discovered before the attack, usually
using lightweight scanning or probing that involves higher layer
protocols.  In contrast, client honeypots actively search for suspicious
entities and solicit interaction with them, such as web crawlers
visiting malicious websites.

Honeypot classification is largely academic.
Since many honeypot variants exist, a distinction is not always possible, nor practical.
In practice, the terms for low- and medium-interaction honeypots are
often used interchangeably.

Methods to distinguish attacks from other types of traffic collected at
honeypots have been proposed.  With the advent of reflective
amplification attacks, server honeypots for the the sole purpose of
capturing DRDoS attacks have been designed, implemented, and deployed.
We discuss amplification honeypot platforms in detail in
\autoref{sec:honeypot:projects}.

\paragraph{Network telescopes} Network
telescopes~\cite{cgcps-omeis-05,azmt-maubp-06,wkbjh-ibrr-10,dkcpp-assb-15,gbd-ripst-16,rb-sssim-19}
are an unsolicited traffic measurement approach that captures incoming
traffic to otherwise unused address space within a larger network
segment.  These typically cover between a \texttt{/8} and \texttt{/24}
of IPv4~address space.  Originally, network telescopes were fully
passive and the network segments were never used to originate any
traffic.  They capture both backscatter traffic (\ie replies to spoofed
addresses of the telescope) and scan traffic.  With the increased
deployment of malicious two-phase scanners~\cite{itd-liuis-21}, \ie
attackers that first check whether a TCP~service is available before
they initiate application requests, reactive telescopes have been
proposed~\cite{hiesgen2022spoki}.  Reactive network telescopes implement
the TCP connection handshake to gain additional knowledge about the
attacker, since the attacker will proceed with an application layer
request.

\subsection{Monitoring Spoofed DDoS Attacks}

When monitoring traffic two crucial questions arise.  \one~Where should
network probes be deployed?  \two~Which packets belong to which type of
event (\eg scan, attack)?

Non-spoofed traffic, or direct-path attacks, can only be observed by
systems that are deployed between the attack source, the destination
target, or at the endpoints.  For example, an appliance to mirror
traffic might be located alongside a victim service, at a network
ingress port, or within an Internet exchange point.  Collecting on-path
observations is a challenge for most researchers and the
ability to capture related but distinct direct-path attacks can be
difficult.  In contrast, reflective attacks allow for broader
observations because they involve triangular packet flows with the
host sending a spoofed packet, a reflector (\eg honeypot) of the
spoofed packet, and the victim host receiving the response to a spoofed
request.

Many reflective amplification attacks rely on amplifier lists to quickly
and successfully conduct attacks.  The lists are commonly curated by
third parties and sold to attackers.  These lists may contain a subset
of all known and currently active amplifiers.  When monitoring
amplification attacks, an amplification honeypot should emulate
amplifier behavior to be appealing to attackers.  To minimize harm,
amplification honeypots typically apply a rate limit to satisfy
amplifier discovery, while avoiding the reflection of meaningful attack
traffic to a victim.

\section{Methodology}
\label{sec:methodology}

We now describe our methodology to systematize, contextualize, and evaluate research about amplification~honeypots.

\subsection{Systematization and Contextualization}
\label{sec:methodology:context}

Our systematization of knowledge aims for an overview and systematic comparison of amplification honeypot research.
This systematization is based solely on previously published work, describing presented methods, data sources, and deployments.
Our framework includes the following parts.

\paragraph{Selecting honeypot research}
We select six honeypot platforms by conducting a systematic literature review searching venues dedicated to security (\ie Oakland, EuroS\&P, Usenix Sec, CCS, NDSS) and measurement (\ie IMC, PAM, TMA) research, as well as broader networking venues (\eg SIGCOMM), covering the last ten years.
The six honeypot platforms and configurations discussed in this paper are seminal for research on amplification attacks.

\paragraph{Comparing honeypot deployments}
We compare honeypot deployments by their setup configuration, \ie number of sensors, duration of deployment, and the geographical as well as topological distribution.
Moreover, we describe which protocols are supported by the honeypots.

\paragraph{Comparing attack inference}
We introduce precise language for describing heuristics that infer attacks from a sequence of packets captured by honeypots.
Then, we show the attack definitions applied by the various honeypot deployments, \ie what are the exact attack thresholds and how are these conveyed in each publication.

\paragraph{Comparing completeness claims}
By considering a realistic attack volume and protocol properties as well as public knowledge about the number of deployed amplifiers, we deduce that attackers can easily impede detection by honeypots.
We show how honeypot research still collectively claims nearly complete attack visibility, despite the lack of ground-truth attack data and the possibility that attackers may hide.

\subsection{Data-driven Evaluation}
\label{sec:methodology:evaluation}

We extend our SoK by conducting a data-driven evaluation.
This is necessary because key methods and assumptions in honeypot research cannot be validated without external observations.
Based on results derived by our contextualization (see \autoref{sec:methodology:context}), we identify further research questions and explore them. 
In detail, \one we analyze whether different attack thresholds used in prior work have a significant effect, \two we verify whether honeypots observe all Internet-wide attacks, and \three we explore the possibilities to improve thresholds.

\paragraph{Evaluating attack thresholds}
We assess the comparability across honeypot projects by describing and analyzing the effects of various flow identifiers and attack thresholds.
To this end, we apply both flow identifier types used in honeypot research.
We then explore the effects of the complete threshold configuration spectrum w.r.t. temporal (\ie timeouts) and volumetric (\ie packet number) properties.
We do this on the dataset obtained by the CCC honeypot platform.

\paragraph{Evaluating attack completeness}
The stability of observations (\emph{honeypot convergence}) is used to justify that honeypot sensors capture a representative view of all Internet-wide attacks (\emph{completeness}).
To validate this, we first review the convergence metric by an optimal, best-case analysis and then proceed with a randomized approach. 
Following this, we check whether the (converging) CCC honeypot platform observes a set of baseline attacks against customers of a leading DDoS mitigation provider.  

\paragraph{Evaluating detection potentials}
We evaluate whether attack detection thresholds can be improved.
We do so by correlating honeypot, telescope, and our baseline data sets.
First, we use the DDoS baseline and try to optimize towards this data set, \ie we improve the honeypot attack detection (but risk over-training towards this specific baseline).
By adopting very permissive thresholds, we infer the upper bound of attack detection.  
Second, we use telescope baseline data to infer whether attack detection thresholds for honeypots already effectively remove baseline scan~events.

\section{Amplification Honeypot Platforms}
\label{sec:honeypot:projects}

We now describe some of the best known honeypot deployments as originally
presented in their publications.  They implement attack detection
mechanisms to identify reflective amplification attacks based on the
packets they receive.  These detection mechanisms, see
\autoref{sec:thresholds}, can be applied on any data but were presented
alongside the data collection platforms described here.

\paragraph{AmpPot} AmpPot~\cite{kramer2015amppot} deploys 21 sensors
supporting nine protocols.  The sensors are primarily deployed in ISP
environments with half located in Japan and the others spread globally.
These sensors are usually configured with static IP addresses, but a
quarter receive dynamic addresses with lease times of up to 51 days.  An
AmpPot sensor can operate in three modes: \one \textit{emulated} runs a
partial, internal implementation of the protocol, \two \textit{proxy}
forwards to a separately deployed service, or \three \textit{agnostic}
amplifies with random data independent of the protocol.

\paragraph{AmpPotMod} AmpPotMod~\cite{noroozian2016amppot} uses a subset of the original AmpPot deployment: eight sensors running in proxy mode (except for SSDP) deployed at ISPs in Japan.
The sensors support up to six amplification protocols and use dynamically assigned IP addresses.

\paragraph{CCC} The Cambridge Cybercrime Center
(\textit{CCC})~\cite{thomas2017thousand} platform is a distributed
honeypot platform that supports eight protocols.  For NTP and DNS, the
sensors proxy to real services. In other cases they respond with a
limited, emulated answer.  The number of sensors fluctuates over time
with a median of 65 active sensors (currently 50).  Sensors are spread
across 10 countries in academic and cloud networks, located in 31 IP
prefixes in 8 ASes.  16 sensors are deployed in their own /28 subnet.
The remaining sensors are deployed at low-cost cloud providers and in a
handful of consumer ISPs.

\paragraph{NewKid} The NewKid platform~\cite{heinrich2021kids} deploys a
single sensor supporting 9 protocols in a university network.  The
sensor operates in proxy mode for Memcached and DNS, and emulates
responses for other services.

\paragraph{HPI} The HPI platform~\cite{griffioen2021adversarial} deploys a total of 549 honeypots %
distributed over five cloud providers and across four continents.
The sensors support six protocols (emulated and proxied) in four different modes that signify the protocol correctness and the amplification factor: \one real-small \two real-large \three fake-small and \four fake-large.

It is worth noting that all platforms deploy a form of rate limiting to minimize adverse effects.
\autoref{tab:honeypot_threshold} summarizes the \emph{setup} properties of the different honeypot platforms.

\paragraph{Impact on other research areas}
The groundwork on amplification honeypots was published in three
consecutive years, AmpPot~\cite{kramer2015amppot} in 2015, AmpPot
Mod~\cite{noroozian2016amppot} in 2016, and
CCC~\cite{thomas2017thousand} in 2017, followed by
HPI~\cite{griffioen2021adversarial} in 2021.  According to Google
Scholar, the oldest honeypot, AmpPot, has been cited the most, reaching
nearly three times the citation count of the others.  With a few
exceptions, all papers are cited in security-related research but have
had influence in multiple, related areas.
The most impactful citations of AmpPot relate to research on technical
aspects of DoS, while AmpPotMod and CCC receive more attention from
adjacent areas such as CRIME-related research.  Measurement research has
more commonly cited AmpPot and CCC compared to AmpPotMod.

\section{Data Sets for Data-driven Evaluation}
\label{sec:data-sets}
We now introduce our data sets, which are summarized in \autoref{tab:data_for_validation}.

\begin{table}%
\setlength{\tabcolsep}{1.8pt}
  \caption{Data sources utilized in this paper to revisit common methods to assess honeypots. All data sources span November~01,~2021--January~31,~2022.}
  \label{tab:data_for_validation}
  \centering
  \begin{tabular}{l c c c}
    \toprule
      Data Source & Attack Thresholds & Convergence & Completeness 
      \\
      & (\autoref{sec:thresholds:comparability}) & (\autoref{sec:convergence}) & (\autoref{sec:completeness})
      \\
    \midrule
      CCC Honeypot Events & \cmark & \cmark & \cmark \\ 
      DoS Mitigation Provider & & & \cmark \\
      US \& EU Telescopes & & & \cmark \\
	\bottomrule
  \end{tabular}
\end{table}
\setlength{\tabcolsep}{5.25pt}

\subsection{Honeypot Data}
We use data from the CCC~honeypot platform.
CCC supplies two types of log formats.
First, a list of victims inferred by applying the default CCC~thresholds. 
Second, a list of all event summaries per sensor.
We analyze the second list for testing various thresholds and validate our scripts with the first list by applying the default CCC~thresholds and inferring the same victims as CCC~did.

We also check whether the CCC platform operated without interruptions.
This eliminates a possibly skewed convergence behavior due to external reasons, \ie a honeypot sensor running only during a fraction of the measurement period would always observe
different attacks than a second sensor running at different times.

\subsection{Telescope Data}
Scanning observations vary between telescopes that differ by topological and geographical properties~\cite{hiesgen2022spoki}.
This is why we use a \texttt{/24} telescope from the US and three \texttt{/24} telescopes from the EU.
In total, 85\% of the CCC honeypot sensors are deployed in these regions, which enables a fair comparison.

Our analysis is based on the assumption that telescopes primarily
observe scan traffic for UDP.  Because network telescopes are fully passive,
scanners do not detect open amplifiers in these networks, which could be
misused in a subsequent attack event.  This means we do not expect
spoofed traffic arriving at the telescope. Moreover, attackers sending
spoofed queries to a telescope would effectively waste their resources
because there is neither reflection nor amplification possible.  This
makes telescopes a suitable vantage point to identify UDP scanners.

In addition to amplification attacks, other UDP (non-scanning) traffic
can be monitored at network telescopes. In 2015, a total of
134~DNS-based amplification attacks have been inferred during a period
of 6~months~\cite{fachkha2015rsdos}.  However, only a handful of these
attacks have been verified and most attacks exhibit properties of
aggressive scanning rather than attacks, \ie the number of targeted
unique dark addresses equals the number of total packets sent.  These
observations might be due to the early stage of detection methods of
amplification attacks, which, at that time, did not account for fast
scanning methods~\cite{durumeric2013zmap}.

The deployment of the protocol QUIC~\cite{RFC-9000} recently changed UDP
traffic properties at telescopes.  Although QUIC runs on top of UDP, it
requires a handshake to initiate connections, making it susceptible to
state-overflow attacks~\cite{nawrocki2021quic}.  This means that we
observe DoS backscatter targeting UDP in addition to TCP~services.
Identifying QUIC backscatter is easy, however, because attacks originate
from the default QUIC port and a specific group of content servers.
Furthermore, they contain fingerprintable data~\cite{nawrocki2021quic}.
Overall, QUIC backscatter does not interfere with our measurements.

Lastly, accidental misconfigurations might lead to UDP traffic at the
telescope.  We argue that such events are rare and unlikely to reach the
ports associated with amplification attacks.  However, we cannot
completely exclude them.

\subsection{DDoS Baseline Data Set}
We collaborate with the world's largest DDoS mitigation equipment provider with a reported global market share of over 20\% in 2020.
We receive partially anonymized attack information under a non-disclosure agreement for popular amplification protocols during our main measurement period.
In total, we are able to observe all reported attacks for the protocols supported by the CCC~honeypots.

The data provided by the mitigation company is based on a DDoS appliance deployed on the direct links between customers and their upstream providers, \ie they are able to observe all external attacks targeting end hosts in the customer networks.
Attack detection is based on observing volumetric peaks and well-known attack vectors to identify anomalous traffic changes. 
It involves customer feedback, which is important for mitigation (traffic scrubbing), since scrubbing could lead to unwanted packet loss in case of false positives. 

Our data set includes start and stop time of an alert, attack type, and flow selection criteria.
For each attack event, we obtain the list of protocols misused,
destination prefixes receiving traffic as observed by the sensor, but
without a detailed breakdown of traffic volumes by target. Although
inferring the specific targets and the impact from attack from this list
is usually not possible it can be utilized for longitudinal validation.
For each attack inferred at the honeypot, we can check whether it is
covered by a mitigation provider attack event and one of its prefixes.
More specifically, the victim is visible as the source of requests at
the honeypot and the destination of attack traffic at the DoS mitigation
sensor.

\paragraph{Quality of the baseline}
To evaluate the precision of thresholds that are used to detect
amplification attacks at honeypots, ground truth data is necessary.
Such data has to be created independent of the honeypots since choosing
one honeypot as a point of reference for multiple honeypot platforms
will lead to ambiguity for two reasons.  First, each honeypot platform
depends on thresholds.  Second, no single configuration can be selected
as the better reference point without attack event verification.
Unfortunately, there is no public source of ground truth data for
DoS-victims and attack events.  Such information is often considered
private and may inflict unexpected cascading effects, \eg a victim might
experience a loss of customers due to a decreased trust in its systems,
or other attackers might be encouraged to launch follow-up attacks on
weakened systems.  Furthermore, a complete view of DoS attacks is
difficult to obtain, because even with large honeypots, attacks often
only use a very small subset of reflection-capable systems.  So although
research-based methods to observe DoS attacks are documented publicly,
their inferred list of victims often remains private or limited due to
vantage point bias.

Companies, such as our data provider, offering DoS traffic mitigation services and equipment are in
a unique position to identify DoS victims.  These mitigation providers
typically operate on the aggregates of traffic paths and relay points
(\ie routers), observing traffic en route rather than having to reside
in an endpoint that may or may not be involved in an attack.  These
aggregate observation points have the advantage of scale, with the
ability to observe and correlate attack events across an array of
covered systems and networks.  Mitigation providers typically have
aggregate traffic sensors deployed at a variety of customer sites.
Anomalous traffic that is detected can be reported, and may eventually
trigger automatic mitigation such as blackholing~\cite{nawrocki2019down}
or traffic scrubbing~\cite{jonker2016protection}.  Although such
mechanisms are also based on heuristics in practice, operational data
based on such mechanisms produces a confirmed set of victims due to its
immediate mitigation actions.  In practice, a detected attack \one
triggers a report that alerts the customer and optionally \two activates
an automatic countermeasure to protect the target from the attack.
False-positives would lead to unhappy and fewer customers, especially
because some mitigation services charge by the volume of traffic
sanitized.  Also, false-negatives would be reported by the customer
(since its service still experiences quality degradation because of DoS
traffic), which ultimately leads to fine-tuning of thresholds and better
detection. %

We call our data baseline for two reasons.
First, during our measurement period, no customer complained about false positives, so we believe that the detection accuracy is very high.
Second, we also believe that this data set provides a representative visibility into attacks because the DDoS mitigation company has a 22\% market share, and its customers are internationally and topologically (small, medium, large networks) distributed.

Given that the events included in our baseline data set are attacks, honeypot platforms claiming complete coverage should be able to detect these events (and maybe~more).

\section{Detecting Attacks}
\label{sec:thresholds}

Attackers unwittingly use amplification honeypots as reflectors to
conduct attacks.  This helps honeypot operators to observe and quantify
attacks.  To distinguish attack packets from scanning and general
Internet background radiation (IBR), honeypots group packets into
"flows" using a flow identifier (id).  \textit{Attack thresholds} then
identify flows that likely belong to an attack.

Flow ids can be created using commonalities among packets such as the
combination of source/destination address and source/destination port
pairs.  Traditional Internet applications minimally use a five-tuple
flow id (IP protocol, address pair, port pair) to group flows, but all
fields in the IP header, UDP header, and abused protocol could be used.
Minimizing the number of flow id fields while correctly classifying all
packets in a group maximizes efficiency.

In a reflective attack, the request packets an attacker sends will
contain a spoofed source address.  The spoofed address becomes the
destination (victim) for amplified response packets.  This is achieved
by handcrafting packets, which requires the attacker to set all fields
to protocol-conforming values.  Attackers may randomize field values
that may vary by operating system or at run-time, such as the IP ID
field or UDP source port, in order to complicate packet classification
at the honeypots.

Among packets to a honeypot the flow-id of a typical UDP-based
amplification attack requires, \one a spoofed source address associated
with a victim, \two a destination IP address of the amplifier (or
honeypot), \three the destination port that maps to the abused protocol
on the amplifier, and finally \four the source port, which can be chosen
freely by the attacker.  Note, carpet bombing attacks, which target IP
prefixes as opposed to a single victim address, may spoof some portion
of the most-significant-bits in a source address in order to randomize
additional bits in the flow-id.

Other fields, such as the IPv4 ID or TTL can similarly be chosen at
random or set to commonly used values to avoid raising suspicion.
Research shows that some botnets use recognizable values for the source
port, TTL, or DNS values~\cite{noroozian2016amppot}.  For example, the
ports 80 and 123 are often found paired with NTP (port 123)
attacks~\cite{czyz2014taming,kramer2015amppot,noroozian2016amppot} and
make up more than 50\% of the attacks together.  Protocol specific
observations show that source port selection differs among
protocols~\cite{griffioen2021adversarial}: attacks using CharGen, QOTD,
RIP, and SSDP exhibit a hard-coded, stable paired port almost
exclusively while NTP and DNS attacks show a larger range of randomized
ports (about 50\%).  Overall, the selected source port in the request
packets of an attack may be useful to track a specific pattern belonging
to an attack entity, but is otherwise unsuitable as a more generic
traffic classifier.

\begin{figure}
  \centering
  \includegraphics[width=\linewidth]{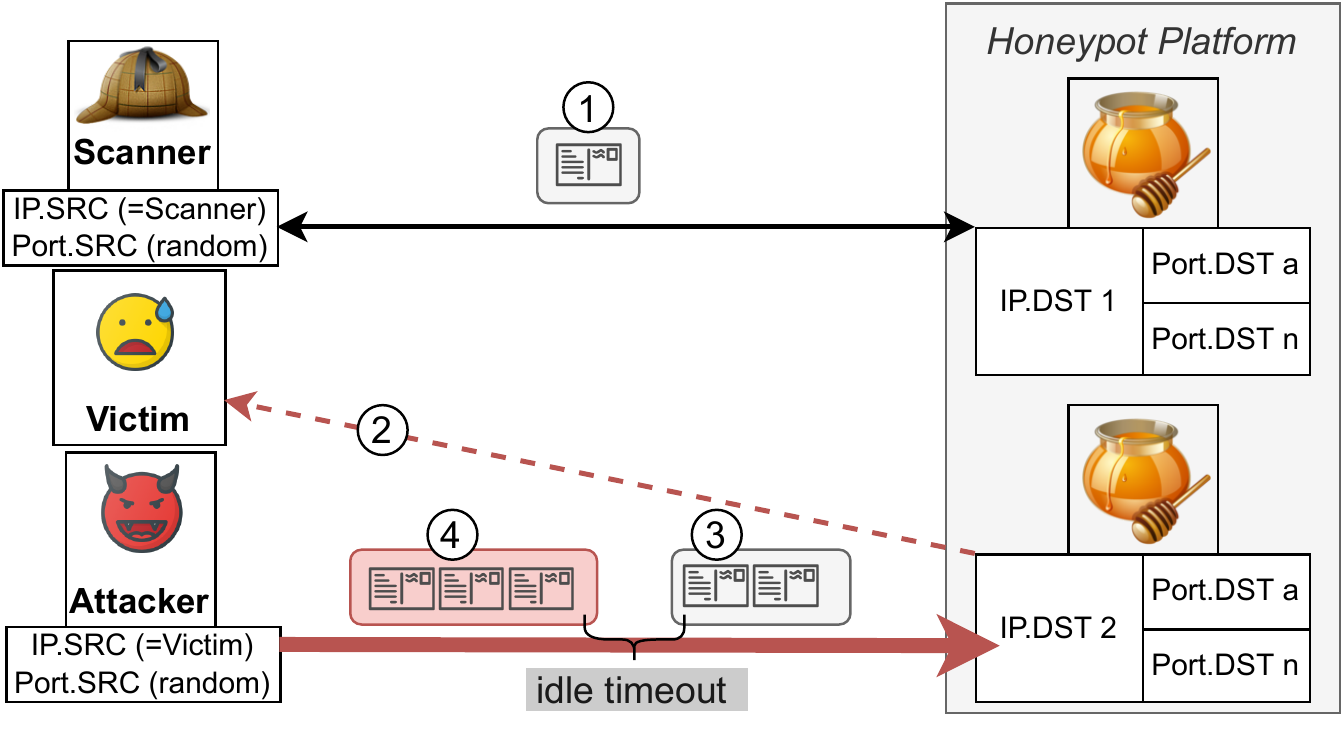}
  \caption{Overview of flow identifiers, timeouts, and packet loads to split a train of packets into flows (rounded rectangles) and attacks (\eg $\ge3$~packets per flow, red rectangle).}
  \label{fig:sketch_overview}
\end{figure}

\begin{table*}[t]
  \caption{Most recent or commonly used amplification honeypot platforms, their setup, definitions of flows, and attack detection thresholds. For CCC, we show the median number of sensors since deployment.}
  \label{tab:honeypot_threshold}
  \centering
  \begin{tabular}{l r c c c c c c c c r r}
    \toprule
     Honeypot Project & \multicolumn{3}{c}{Setup} & \multicolumn{6}{c}{Flow Identifier}  & \multicolumn{2}{c}{Attack Thresholds} \\
       \cmidrule(r){2-4} \cmidrule{5-10} \cmidrule(l){11-12}
      & Sensors & \multicolumn{2}{c}{Distributed} & \multicolumn{2}{c}{\texttt{IP Prefix}} & \multicolumn{2}{c}{\texttt{IP Address}} & \multicolumn{2}{c}{\texttt{Port}} &  Idle Timeout & \makecell[c]{Packet Load}  \\
      \cmidrule(lr){3-4} \cmidrule(lr){5-6} \cmidrule(lr){7-8} \cmidrule(lr){9-10}
      & \makecell[c]{[\#]} & Geo & Topo & \texttt{Src} & \texttt{Dst} & \texttt{Src} & \texttt{Dst} & \texttt{Src} & \texttt{Dst} & \makecell[c]{[minutes]} & \makecell[c]{[packets]} \\
    \midrule
     AmpPot~\cite{kramer2015amppot} & 21 & \cmark & \cmark & \xmark & \xmark & \cmark & \xmark & \xmark & \cmark & 60 & $\geq100$  \\
     AmpPotMod~\cite{noroozian2016amppot} & 8 & \xmark & \cmark & \xmark & \xmark & \cmark & \xmark & \xmark & \cmark & 10 & $\geq100$  \\
     CCC~\cite{thomas2017thousand} & 65 & \cmark & \cmark & \xmark & \xmark & \cmark & \cmark & \xmark & \cmark & 15 & $\geq5$  \\
     NewKid Mono~\cite{heinrich2021kids} & 1 & \xmark & \xmark & \cmark & \xmark & \xmark & \cmark & \xmark & \cmark & 1 & $\geq5$  \\
     NewKid Multi~\cite{heinrich2021kids} & 1 & \xmark & \xmark & \cmark & \xmark & \xmark & \cmark & \xmark & \xmark & 1 & $ \geq2$ ports \& $\geq5$ \\
     HPI~\cite{griffioen2021adversarial} & 549 & \cmark & \cmark & \xmark & \xmark & \cmark & \cmark & \xmark & \cmark & 1 & $\geq2$ honeypots \& $>20$ \\
	\bottomrule
  \end{tabular}
\end{table*}

\begin{table*}
  \setlength{\tabcolsep}{3.8pt}
  \caption{Expected outcome of different attack detection methods, in case of a uniform amplifier utilization and an attack load of 1~Gbit/s lasting 5 minutes.}
  \label{tab:honeypot_visibility}
  \centering
  \begin{tabular}{r l r r r r r c c c c c}
    \toprule
      \multicolumn{7}{c}{Attack Configuration} & \multicolumn{4}{c}{Attack Detected} \\
      \cmidrule(lr){1-7}
      \cmidrule(lr){8-11}
      UDP~Port & Protocol & $\sim$Request Size & Ampl. Factor & \# Amplifiers & Reqs/Attack &  Reqs/Amplifier & AmpPot(Mod) & CCC & NewKid & HPI \\
    \midrule
      17 & QOTD & 15 Bytes & 140 & 31k & 17.9M & 576 & \cmark & \cmark & \cmark & \cmark \\
      19 & CharGen & 15 Bytes & 356 & 30k & 7.0M & 234 & \cmark & \cmark & \cmark & \cmark \\
      53 & DNS & 37 Bytes & 41 & 1.9M & 24.7M & 13 & \cmark & \cmark & \cmark & \xmark \\
      123 & NTP & 13 Bytes & 557 & 2.3M & 5.2M & 2 & \xmark & \xmark & \xmark & \xmark \\
 	  389 & LDAP  & 52 Bytes & 63 & 8k & 11.4M & 1430 & \cmark & \cmark & \cmark & \cmark \\
      1900 & SSDP & 90 Bytes & 31 & 1.9M & 13.4M & 7 & \xmark & \cmark & \cmark & \xmark \\
	\bottomrule
  \end{tabular}
\end{table*}

\autoref{fig:sketch_overview} puts the flow identifier into context.  A
honeypot platform is built from multiple sensors that receive packets
from a variety of sources such as scanners \circled{1}.  The goal is to
identify packets that are not just information gathering but used to
attack victims \circled{2} via reflection attacks.
Packets in the same flow-id can then be grouped together based on an
\textit{idle timeout}, which determines the maximum interval between two
packets belonging to the same flow~\circled{3} or to a different
flow~\circled{4}.  Finally, only flows that contain a minimum
\textit{packet load} are considered \textit{attack~flows}~\circled{4}.

We now introduce the various attack definitions from related work.  See
\textit{Attack Thresholds} in \autoref{tab:honeypot_threshold}.  Note,
attack definitions are independent from the deployments described in
\autoref{sec:honeypot:projects}.  Data from any deployment can be
combined with any attack-detection method.  However, we use the names of
the original publications to distinguish them.

\subsection{Current Methods}
\label{sec:thresholds:current_methods}

CCC considers a flow as an attack flow if it contains at least five
packets \textit{per sensor} within an idle timeout period of 900
seconds.  This is in contrast to AmpPot, which defines higher
thresholds: An attack flow must contain at least 100~packets with an
idle timeout of 3600~seconds or 600~seconds when observed across
\emph{all} sensors.  CCC and AmpPot use the source address and the
destination port to assign a flow id to multiple packets.  CCC also
considers the destination address, \ie the sensor, as an additional
restriction to classify packets into an attack flow.

NewKid describes two types of attacks, monoprotocol and multiprotocol
attacks.  We use the labels NewKid Mono and NewKid Multi to distinguish
them.  Mono requires five packets in an attack flow with an idle timeout
of 60~seconds.  The Multi variant extends this definition to include
packets that have at least two different destination ports within the
idle timeout period.  To account for carpet bombing attacks, \ie attacks
hitting multiple addresses in the same IP~prefix, the flow id uses the
source IP prefix, instead of the address, combined with the destination
IP~address and, for Mono, the destination port.  CCC is also able to
infer carpet bombing attacks but only if 16~\emph{individual} attack
flows target victims in the same \texttt{/24} prefix.

HPI applies an idle timeout of one minute, a packet load of at least 20
packets, and requires activity observed by at least two sensors.
Although their flow id is defined per-sensor, they require at least two
overlapping flows.

\paragraph{How to (not) present thresholds} We find a recurring pattern
that attack thresholds are insufficiently justified.  We acknowledge
that rigorous thresholds are hard to identify without ground truth.
Unfortunately, there is little to no discussion on the effects of the
chosen thresholds.  Documenting its effects is possible without ground
truth and certainly would help the reader in future research.

The AmpPot paper includes a definition paragraph, specifying the minimum
flow filter threshold, stating \textit{sources [must send] at least 100
consecutive requests to our honeypots}~\cite{kramer2015amppot}.  The
authors claim that this is a \textit{conservative} threshold but do not
provide further details on the reasoning or the number of events this
configuration excludes.  We believe it is based on their analysis of
telescope traffic and the behavior of large-scale scanners contacting at
least 64 dark addresses on the same port.  They find that roughly 94\%
of the scanners send less than two packets per IP address on average.

\begin{figure*}%
  \begin{subfigure}{.5\textwidth}
    \centering
    \includegraphics[width=1.0\linewidth]{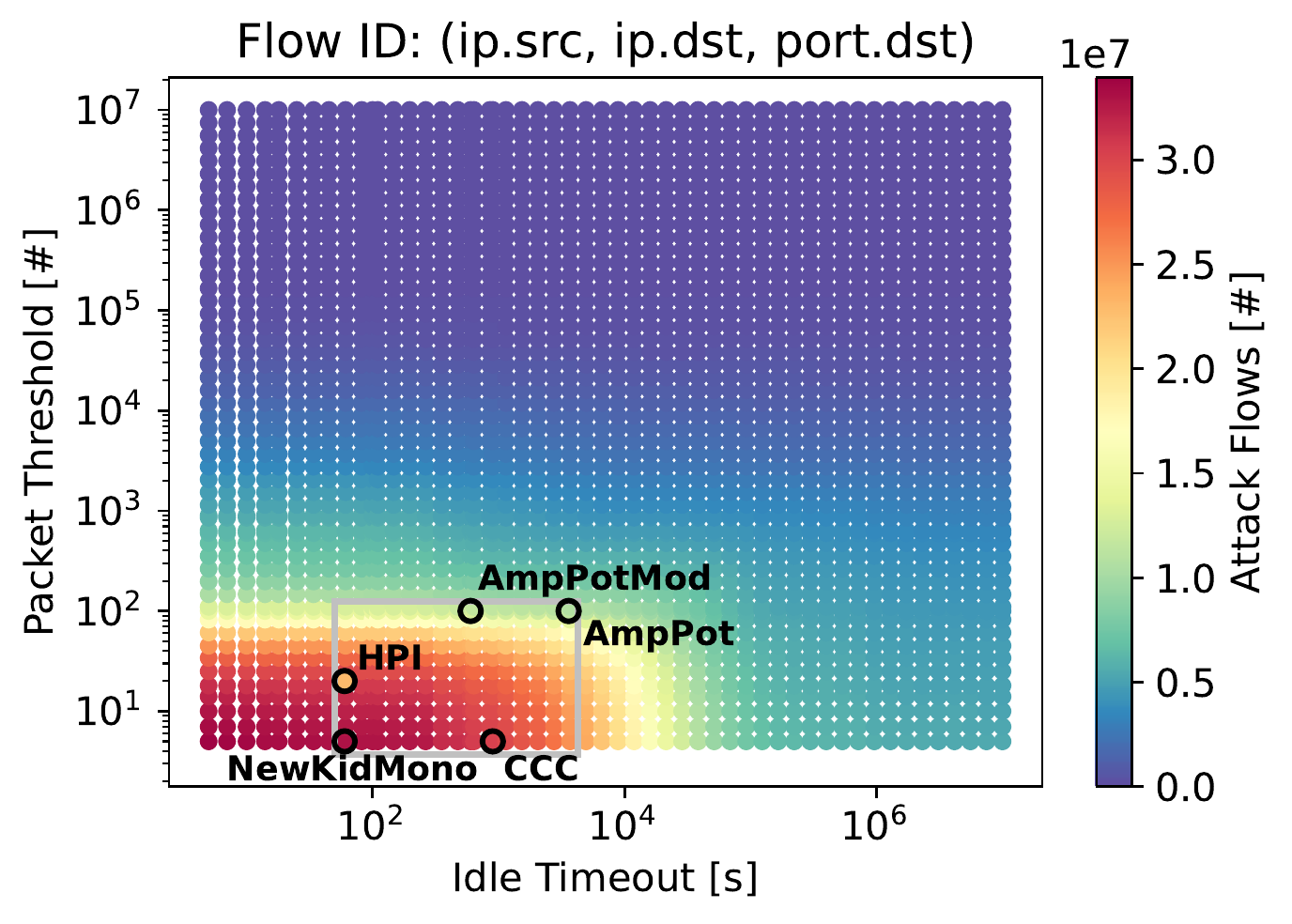}
    \caption{CCC flow identifier: per sensor.}
    \label{fig:threshold:num_attacks_ccc}
  \end{subfigure}\hfill
  \begin{subfigure}{.5\textwidth}
    \centering
    \includegraphics[width=1.0\linewidth]{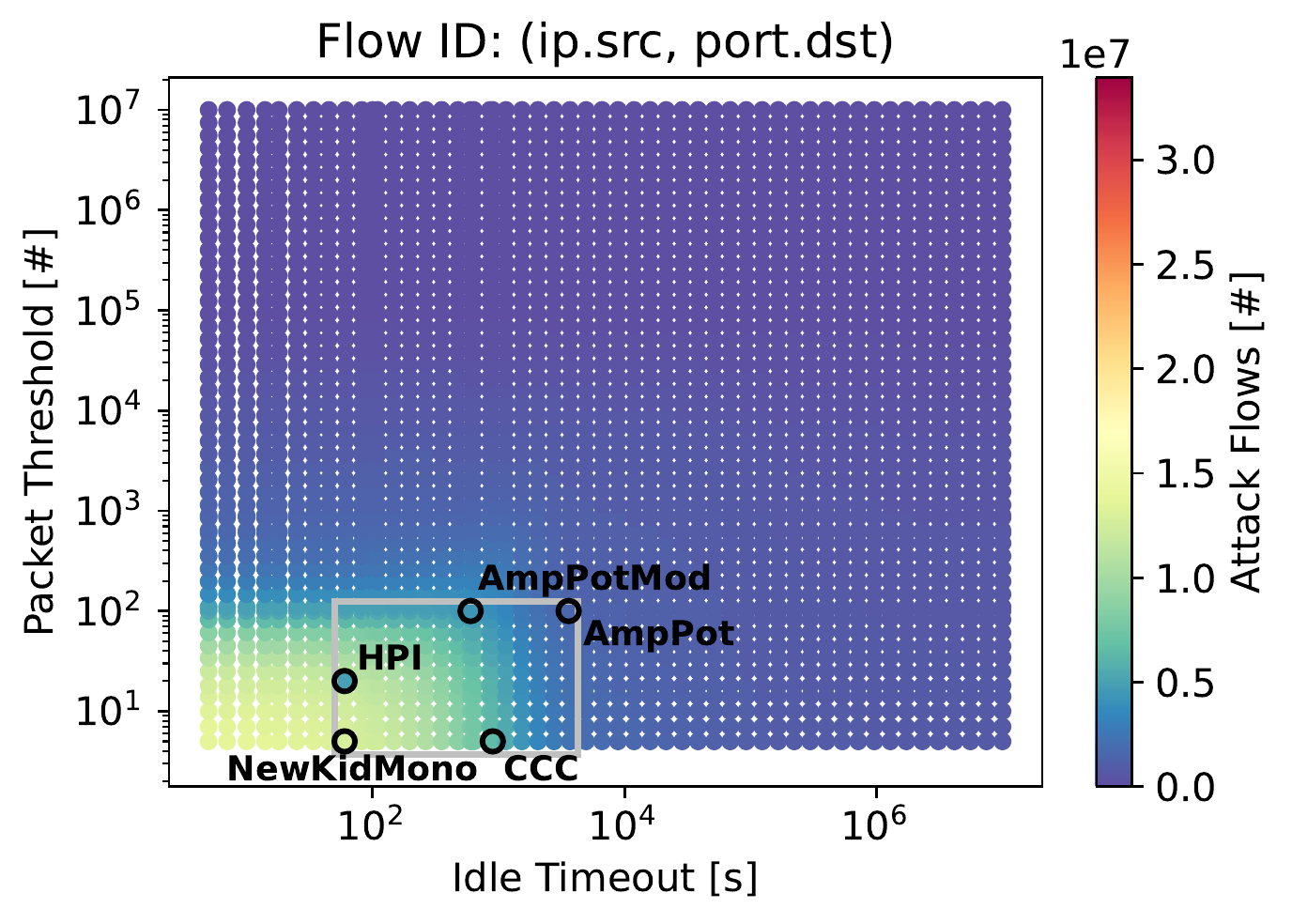}
    \caption{AmpPotMod flow identifier: per platform.}
    \label{fig:threshold:num_attacks_amppot}
  \end{subfigure}\hfill
  \caption{Number of attack flows, depending on different definitions of flow identifiers and attack~thresholds. Thresholds from honeypot research~(\autoref{tab:honeypot_threshold}) are located in the gray box, the HPI~attack threshold marks fewer flows as attack.}
  \label{fig:threshold:num_attacks}
\end{figure*}

In AmpPotMod, the authors reduced the idle timeout to \textit{analyze
attack duration with a more fine-grained
approach}~\cite{noroozian2016amppot}.  It remains unclear how this
change affected their results, \eg the number of detected attacks.

CCC~\cite{thomas2017thousand} selected their idle timeout \textit{to
loosely correspond with the availability of short lived attacks (under
an hour) from booter systems}.  However, they do not provide an analysis
to validate their choice of threshold.

The NewKid paper illustrates that \textit{thresholds were established
empirically}~\cite{heinrich2021kids} by manually analyzing three weeks
of traffic.  The authors infer three traffic classes (\emph{slow},
\emph{fast}, \emph{bursty}) and try to classify the first class as
scanner and the remaining classes as attacks.  The description lacks
detail on this inference and the \textit{automatic classification} in
particular.  It remains unclear how the victim CIDR blocks are selected.

The HPI team states that they \textit{experimentally derived that actors
use up to 20 packets from the same source IP
address}~\cite{griffioen2021adversarial}, but no further explanation is
given about the experiment setup.

All papers include basic reasoning of the chosen attack thresholds.
While the adjustable parameters are similar, the reasoning for different
choices of flow id, packet load, and idle timeout remain unclear in
practice.  We highly encourage future work to use appendices to provide
a more detailed analysis.  This will enable the community to reproduce
data selection processes and inferences.

\subsection{Comparability of Attack Thresholds}
\label{sec:thresholds:comparability}

We now systematically analyze the effects of various flow identifiers
and attack thresholds to assess the comparability across research
projects.  We distinguish between sensor-based and platform-based
flow-identifiers.  Although we include all threshold configurations from
related work, we will focus on the CCC and AmpPotMod configurations.
Their publications have a wide reach and they differ in a key aspect:
the CCC flow identifier is applied per-sensor whereas AmpPotMod is
applied per-platform.
Please note that we do not use ground truth data but rather explore the
effects of the configuration spectrum.  The dataset contains packets
obtained by the CCC honeypot platform.

\begin{figure*}
  \begin{subfigure}{.5\textwidth}
    \centering
    \includegraphics[width=\linewidth]{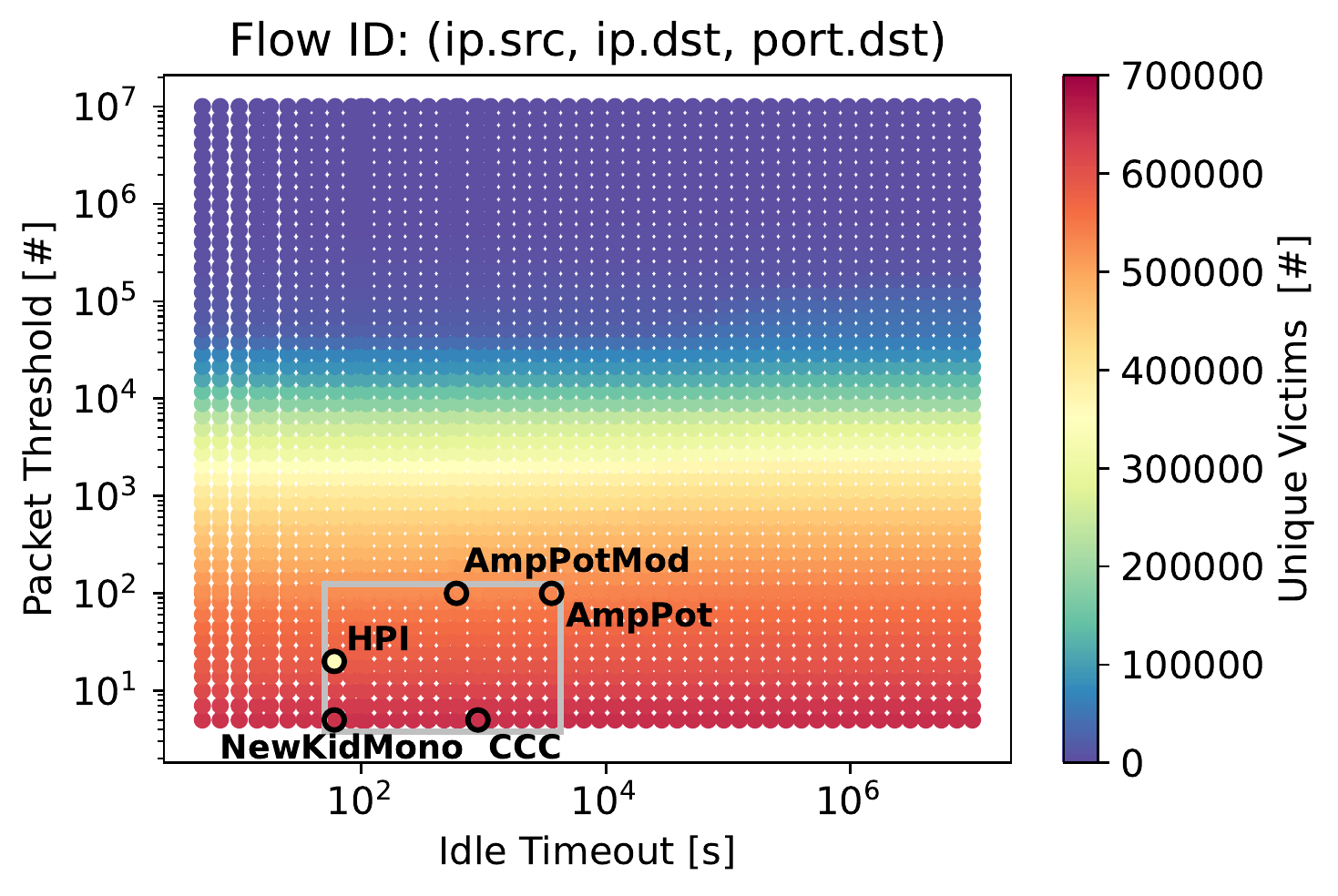}
    \caption{CCC flow identifier: per sensor.}
    \label{fig:threshold:num_victims_ccc}
  \end{subfigure}\hfill
  \begin{subfigure}{.5\textwidth}
    \centering
    \includegraphics[width=\linewidth]{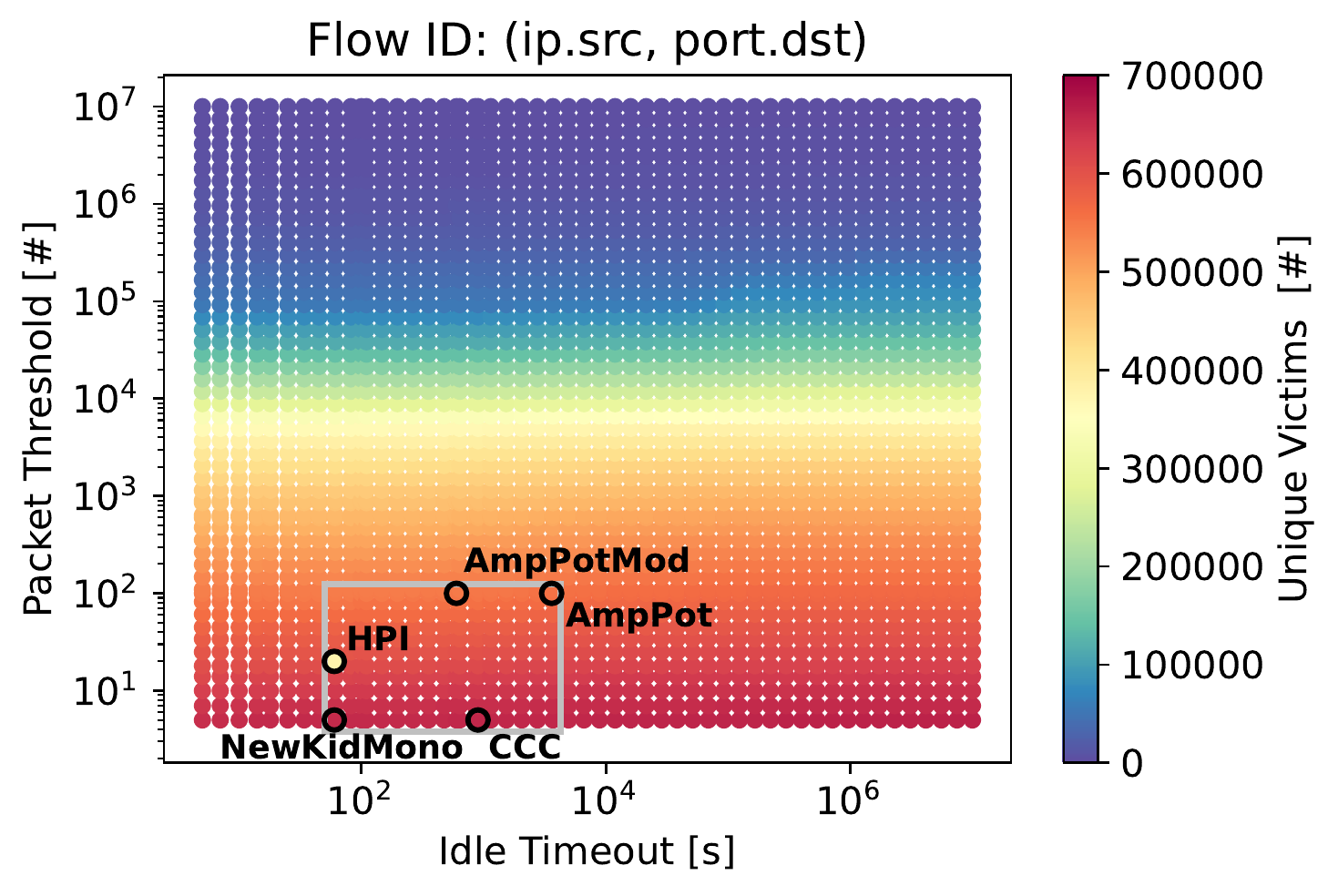}
    \caption{AmpPotMod flow identifier: per platform.}
    \label{fig:threshold:num_victims_amppot}
  \end{subfigure}\hfill
  \caption{Number of victims, depending on different definitions of flow identifiers and attack~thresholds. Thresholds from honeypot research~(\autoref{tab:honeypot_threshold}) are located in the gray box, HPI~attack threshold marks less hosts as victims.}
  \label{fig:threshold:num_victims}
\end{figure*}

\paragraph{Counting attack flows misleads} First, we show the number of
attack flows for different thresholds, see
\autoref{fig:threshold:num_attacks}.  The heat map shows the number of
identified flows on the z-axis as a function of the idle timeout in
seconds (x-axis) and the packet threshold (y-axis).  The maximum x-axis
value is around $10^7$ which correlates to the complete measurement
period of the dataset.  A grey square marks the area for the thresholds
listed in \autoref{tab:honeypot_threshold}.  The left figure uses the
CCC flow identifier, \ie source address, destination address, and the
destination port applied per sensor, whereas the right figure uses the
AmpPot flow identifier, \ie source address and destination port applied
across the whole platform.  For NewKid, we only show the Mono variant
because it was predominantly used in the paper.  The value for the HPI
thresholds is visually striking, because we additionally include the
requirement of at least 2 honeypots sensors for this data point.

We infer two findings: \one The platform-based flow identifier counts
less attack flows because it groups packets across different sensors
into the same flow instead of counting the flows per sensor -- provided
attacks utilize multiple sensors.  We find 12.9M attack flows with
AmpPotMod thresholds and 30.3M with CCC thresholds when applied to the
per-sensor flow ids (\autoref{fig:threshold:num_attacks_ccc}), and 4.4M
and 6.4M attack flows when applied to the per-platform flow ids,
respectively (\autoref{fig:threshold:num_attacks_amppot}).  \two Longer
idle timeouts only affect the attack flow count up to $\sim10^4$ seconds
(3 hours), but have negligible effect thereafter.  At that point short
consecutive attacks are likely grouped into a single flow.
The idle timeout has a stronger effect on the per-sensor flow identifier
because it is less likely to observe packets at the same sensor.

\paragraph{Detected victims uncover high similarity}
We now analyze the number of detected victims, see
\autoref{fig:threshold:num_victims}.  The figure uses the same $x$ and
$y$-axis as \autoref{fig:threshold:num_attacks} but shows the unique
victim count on the $z$-axis (the maximum is two orders of magnitude
lower).  Instead of counting attacks or attack flows---which are heavily
influenced by the choice of flow identifier: per-sensor vs
per-platform---we count the number of victims.  Since both approaches
are run on the same data, measurements are comparable.  Note,
that the number of victims should be a lower bound of the attack numbers.
The idle timeout still affects results as a long idle timeout might
group packets from low volume scanners into attack events, thus
potentially generating victim artifacts.

For the per-sensor flow identifier, we find 644k victims using the CCC
threshold and 531k victims using the AmpPotMod threshold.  For the
per-platform flow identifier, we find 654k and 549k victims,
respectively.  By comparing the respective configurations (CCC flow
identifier and CCC thresholds versus AmpPotMod flow identifier and
AmpPotMod thresholds) we find only a difference of 15\%.  Visually, both
configurations are present in the same cluster and gradient.
Reassuringly, this means that the results of the various honeypot
platforms are indeed comparable.  An exception to this finding are the
HPI thresholds, which require at least two sensors to observe attacks. 
This leads to a 45\% smaller victim set.

\subsection{Evading Threshold-based Detection}
\label{sec:ecosystem:protocols}

Current studies (see \autoref{sec:honeypot:projects}) apply a single threshold configuration that is independent of the misused protocol.
The CCC~honeypot detects NTP~(60\%), LDAP~(31\%), and DNS~(4\%) as the most popular amplification protocols in 2022.
This observation confirms common expectations, which assume attackers choose protocols that allow for high amplification and provide a rich amplifier infrastructure.
NTP, for example, does not only provide the highest amplification factor and many amplifiers but also \textit{mega-amplifiers}~\cite{czyz2014taming}, \ie hosts that exhibit a significantly larger amplification factor due to their configuration, making this protocol most appealing to attackers.

The use of a protocol-independent threshold is surprising, though, since each protocol exhibits
features (\ie amplification) and deployment (\ie instances in
the wild) that may be leveraged  by attackers in different ways.
We now analyze
the attacker potentials to impede detection by honeypots. We do not use honeypot
measurements but model a realistic attack volume and utilize protocol
properties as well as public knowledge about the number of deployed
amplifiers---similar to what attackers can do.

\looseness=-1
We assume a simple attacker model: attackers try to minimize exposure by
reducing the load per amplifier while still achieving a desired traffic load. Given a set
of amplifiers and a target attack load, an attacker uniformly distributes
connection requests among all amplifiers.  Based on this model, we infer the
number of expected packets per amplifier for an attack load of 1 Gbit/s lasting
5 minutes.  This attack scenario is realistic and produces more traffic than
the majority of attacks: \one Although new attack traffic peaks are reached
yearly, the majority of attacks (98\%) do not exceed 1 Gbit/s, even in the year
2021~\cite{cloudflare2021radar, netscout2019report, kaspersky2019report}.  \two
A recent honeypot platform observes that 50\%--80\% of amplification attacks
are shorter than 5 minutes \cite{griffioen2021adversarial}, depending on the
protocol.
Triggering high volume attacks by requesting relatively little from a large number of amplifiers is doable given current amplification factors and deployment of amplifiers (see, \eg NTP or SSDP).
We adopt amplification factors from related work \cite{r-ahrnp-14},
the numbers of open amplifiers from publicly accessible scan
projects~\cite{shadowserver2022website, shodan2014website}, and then apply
common attack detection thresholds.

\autoref{tab:honeypot_visibility} lists the calculated attack
configurations and compares them against the attack thresholds presented
in the AmpPotMod, CCC, NewKid, and HPI papers.  Depending on the amplification
protocol each honeypot sensor would experience different packet loads,
ranging from 2~(NTP) to 1430 (LDAP) packets during the attack time.
Attacks that require fewer requests per amplifier tend to remain unnoticed by current detection methods.
This result highlights that current detection methods may miss smartly tailored attacks and that thresholds can best detect attacks when the packet load per amplifier is high.
Overall, this suggests that the honeypot observations are incomplete.

We conclude that honeypot observations cannot be simply explained \emph{in situ} but have to be embedded into the protocol ecosystem and the decision-making that determines amplifier lists used by attackers.

\section{Honeypot Convergence}
\label{sec:convergence}

In this section, we revisit  accuracy estimations for observations from a distributed honeypot
platform.  We explore the notion of \emph{honeypot convergence}, a completeness 
measure of the detections that is  influenced by the number of honeypots deployed and their
configuration.  We evaluate the impact of varying deployment scenarios
based on the CCC platform.

\subsection{Current Methods}
\label{sec:convergence:current_methods}

Honeypot convergence is based on the assumption that the observed event set stabilizes (\ie converges) as the set of honeypot probes varies.
It is considered a key property of a honeypot platform, because it provides a comparative measure for attacks observed by different honeypot deployments.
Convergence \emph{supposedly} occurs when a set of honeypot probes provide a complete view of all attack events.

\looseness=-1
In the AmpPot paper~\cite{kramer2015amppot}, the authors order all
the honeypot probes by name and then compute the running sum of new
attacks contributed by each probe in turn.  They conclude that
$10$~AmpPot probes identify $>$ 90\% of all attacks and that additional probes
add only very few new attacks.

\begin{figure*}
  \begin{subfigure}{.5\textwidth}
    \centering
    \includegraphics[width=\linewidth]{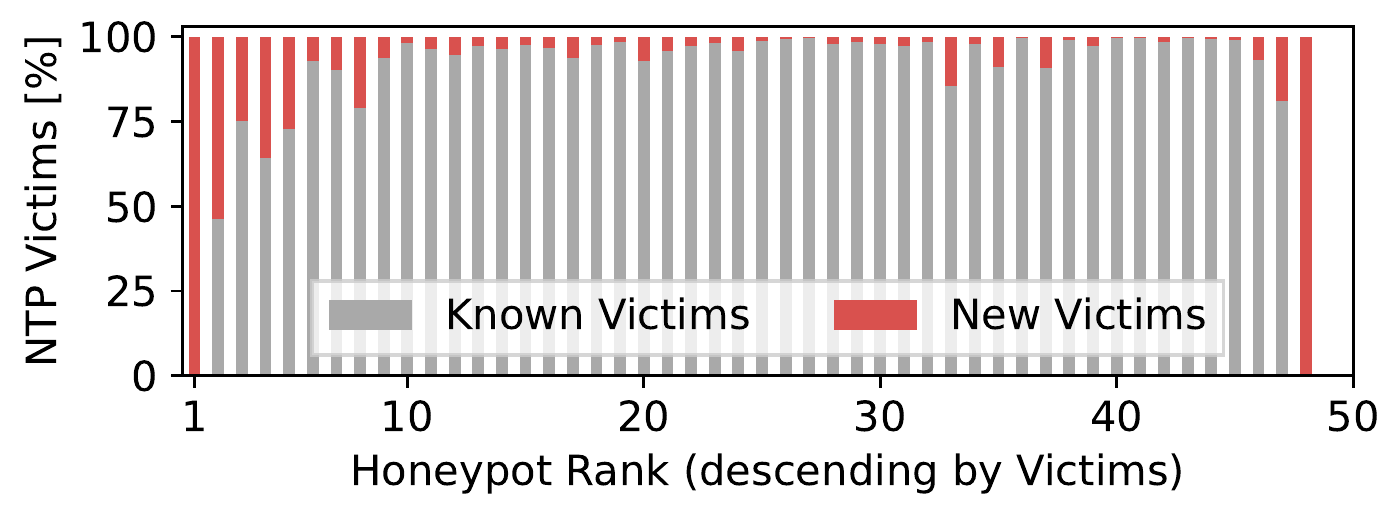}
    \caption{New victims per sensor.}
    \label{fig:precision:convergence_ntp_sbars}
  \end{subfigure}\hfill
  \begin{subfigure}{.5\textwidth}
    \centering
    \includegraphics[width=\linewidth]{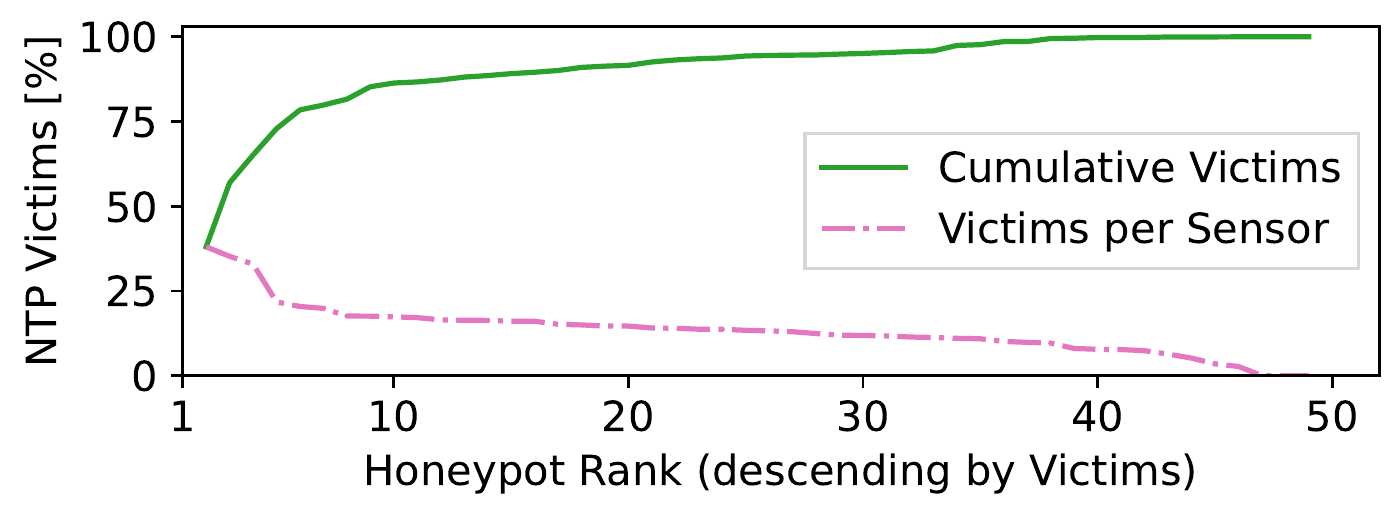}
    \caption{Victims per sensor and cumulative total.}
    \label{fig:precision:convergence_ntp_cumsum}
  \end{subfigure}\hfill
  \caption{Convergence behavior for NTP using a near-optimal selection of honeypot sensors.}
  \label{fig:precision:convergence_ntp}
\end{figure*}

In the CCC~paper~\cite{thomas2017thousand}, the authors apply a
capture-recapture analysis, a statistical method known from ecology,
which derives the number of estimated attacks by random sampling of the
honeypot probes.  They conclude that the CCC platform captures 85.1\%--96.6\% of all attacks.
Other work derives that already 5~CCC~sensors converge and monitor $>$ 99.5\% of the DNS victims~\cite{nawrocki2021ecosystem}.

Although the number of sensors is significantly higher ($\sim$150),  the authors of the HPI deployment  claim to have a complete view also on the basis of convergence behavior. %

The stabilization of attack events (\ie convergence) when adding more
probes is a common justification for specific honeypot settings.
It remains premature, though, to conclude from convergence that a complete set of attack events has been observed. Convergence also occurs if a large set of attack events never enters the honeypot platform. 
  Recent research observes this for 
different honeypot deployments, which show  very diverging event sets with 
incomplete pictures of  attacks.  Two independent studies show small overlaps of
only 4\%~\cite{nawrocki2021ecosystem} and  8.18\%~\cite{kopp2021ddos}
between UDP amplification attacks observed at common honeypots and different
vantage points (\ie other honeypot platforms and IXPs), challenging previous
assumptions and claims of convergence.
Furthermore, analyses based on the large HPI platform show that convergence
differs by protocol and that a general approach to high attack visibility (\ie 99\%) is hard to
achieve, \eg RIP~measurements require 60~sensors and other
protocols $\approx$150~sensors~\cite{griffioen2021adversarial}.

\begin{figure*}
  \begin{subfigure}{.49\textwidth}
    \centering
    \includegraphics[width=\linewidth]{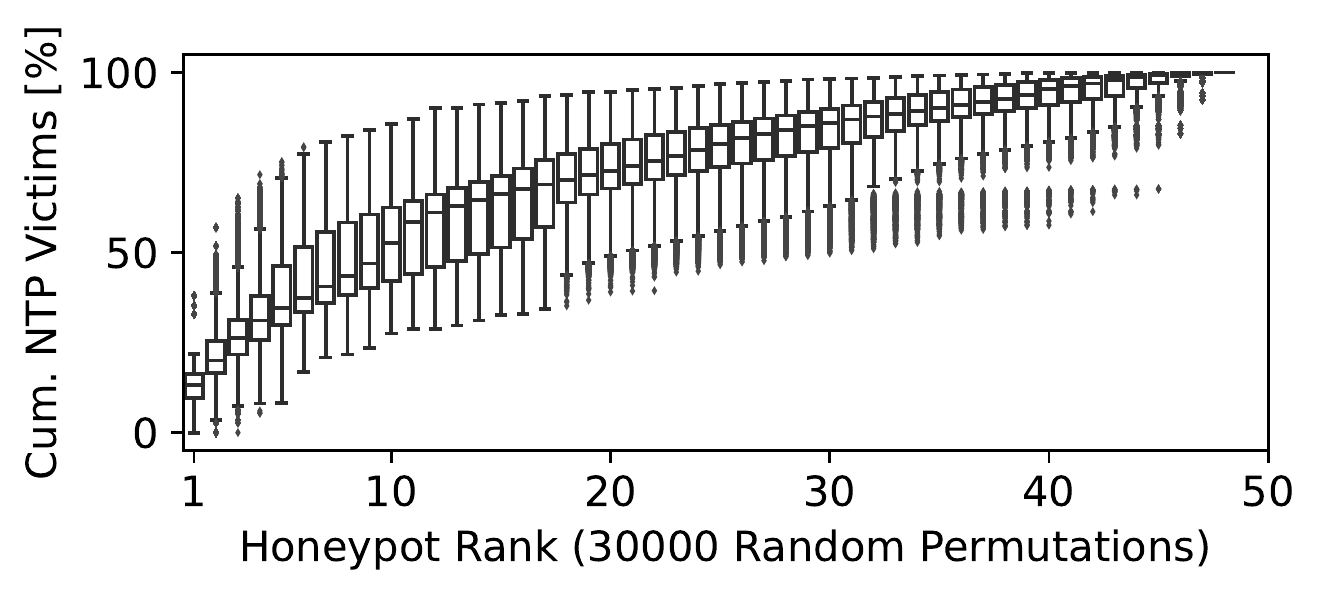}
    \caption{Convergence: High variances in the results suggest that convergence is less stable than previously assumed.}
    \label{fig:precision:permutations_ntp}
  \end{subfigure}\hfill 
  \begin{subfigure}{.49\textwidth}
    \centering
    \includegraphics[width=\linewidth]{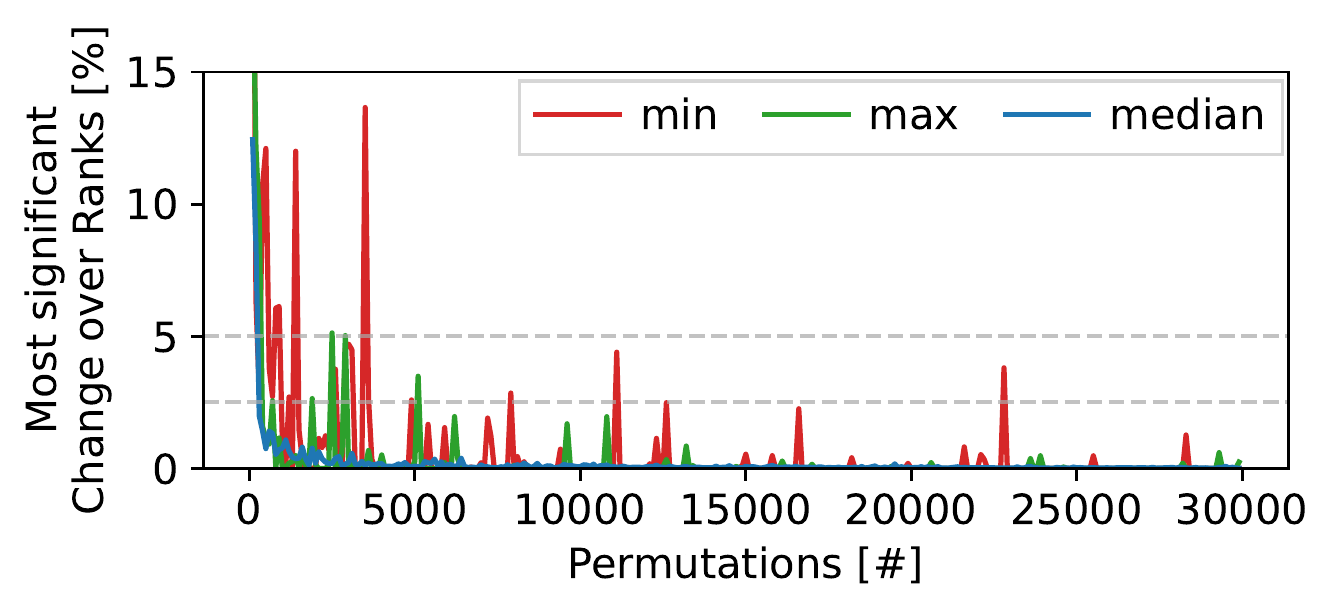}
    \caption{Relative differences of min, med, and max of detected victims. Even at $\approx$25k, worst-case results (min) differ by less than 2\%.}
    \label{fig:precision:nperm_diff_ntp}
  \end{subfigure}\hfill
  \caption{Examining the convergence of NTP over 30k permutations.}
  \label{fig:precision:ntp}
\end{figure*}

Reviewing the implications of honeypot convergence is important because
this measure has been used as a fundamental building block for the
justification of honeypot results. Given the visibility of a honeypot platform, 
researchers had no other means but to test for the convergence of their results. 
We argue, however, that honeypot convergence should be re-interpreted, as it is only a fair measure of the limits of visibility, \ie a test whether the horizon of the platform has stabilized. 

\subsection{Reproducing Convergence}
\label{sec:convergence:reproducing}

We use data from the CCC~honeypots (see \autoref{sec:thresholds}) to
illustrate that the strategy of selecting probes has a significant
impact on convergence results.
Using the default CCC thresholds, we learn about 1.4M~attacks towards 644k victims
during our measurement period spanning 3~months.  The most common
protocols for amplifications are NTP~(60\%), LDAP~(31\%), and DNS~(4\%).
We observe continuous scans or attacks for all but one faulty
sensor for NTP and LDAP,  why we conclude that these
services were run throughout the whole measurement~period.

We now reproduce the honeypot convergence based on a near-optimal sensor
selection, analogous to prior
work~\cite{kramer2015amppot,nawrocki2021ecosystem}.
We sort the sensors by the number of victims and perform a greedy selection, \ie we select the sensors with the most unique victims first.
\autoref{fig:precision:convergence_ntp} exhibits the results for NTP, for LDAP we refer to Appendix~\ref{apx:ldap_convergence}. The share of new victims, which an individual sensor
contributes, decreases quickly until rank~10.  For NTP at rank~10, 87\%
of victims have been already observed and the subsequent sensors do not
significantly increase the cumulative count although each sensor
observes $\sim$16.5\% of all victims.  For LDAP at rank~10, we observe
slightly fewer victims (76\%).  Each additional sensor
observes $\sim$35\% of all victims but increases the cumulative share only by
 $0.5$\%.
In summary, we successfully reproduced the honeypot convergence for the given 
platform and measurement period.

\subsection{A Fair Convergence Introspection}
\label{sec:convergence:fair_introspection}

This convergence measure, which we just reproduced, follows a probe sampling that 
prefers sensors with a large number of common victims. As such, it 
 is biased towards fast convergence.   
We now want to analyze the general convergence behavior and answer the question
 whether this bias leads to missing relevant data from the result set.

In general, the convergence behavior depends on the number \emph{and the order} of
considered sensors.
To eliminate order bias, we create 30k~random permutations of all CCC~sensors and
re-inspect convergence for NTP, see
\autoref{fig:precision:permutations_ntp}. This analysis differs from
related work~\cite{griffioen2021adversarial} by exploring further 
statistical details instead of only averages.  Each box includes
the median (bar), up to $1.5\times$ of the interquartile range
(whiskers), and all minimal and maximum values (outliers).  This plot
clearly visualizes the large variances across convergence results, depending
on the combination of probes.  Considering the best (upper outliers) and
worst (lower outliers) case scenarios of 20~sensors (rank 20), we find
39\%--95\% of NTP victims.  Furthermore,
the upper outliers resemble very closely the cumulative ratio of victims
in \autoref{fig:precision:convergence_ntp_cumsum}.
These observations lead to two insights.
Fist, they confirm our previous
observation that probes with higher weight (\ie more attacks) introduce
a bias towards fast convergence.  Second, they emphasize that
convergence measures should be utilized with great caution when 
justifying the completeness captured by honeypot deployments.

We still want to justify that 
 we do not compute \emph{all} permutations of
currently 50~CCC~sensors due to numerical complexity ($50! \approx3\cdot10^{64}$~permutations).
Limiting to 30k~permutations already shows stable results.  To assess
the stability, we iteratively create 100~new permutations and add them
to the total set of permutations.  For each set of permutations, we
determine the largest differences of the minima, median, and maxima of
detected victims across all ranks.  The results are shown in
\autoref{fig:precision:nperm_diff_ntp}.  After an initial phase of
significant changes, the median becomes very stable using at least
25k~permutations.  Occasionally, minima can change up to $\sim$2\%, 
\textit{cf.} Appendix~\ref{apx:ldap_convergence} for LDAP.

It is noteworthy that the capture-recapture method can be inadequate for estimating an unknown population.
Related work finds that
\one accuracy depends on capturing a large proportion of the population~\cite{lee2014catch}, \ie the majority of attacks, and
\two it looses accuracy for transient populations~\cite{tilling1999capture}, \ie when attackers cease or move between measurement areas due to new amplifier lists.
All this makes it very likely that the estimated number of attacks accounts only for a subset of total attacks.

\subsection{Convergence versus Completeness Metrics}
\label{sec:convergence:vs_completeness}

In the previous sections, we have shown that convergence is not a stable metric but (if cautiously applied) can shed light on the horizon of visibility for a honeypot platform deployment. The completeness of the observation (\ie the detection of all ongoing attacks), however, strongly depends on how an attacker selects the amplifiers. Consider two corner cases and one likely scenario.

\begin{enumerate}
	\item An attacker may not select at all but send spoofed requests to arbitrary IP addresses. In this case, the probability of observing the attacker is extremely low for any given honeypot platform.

	\item An attacker may---after scanning---use all amplifiers of a given protocol. In this case, a single sensor suffices for detecting the attack.

	\item An attacker may use a limited subset of amplifiers, \eg an amplifier hit list. This list may have been collected according to efficiency (\ie amplification factors), (geographic or topological) locality, or other means. In this case, the probability of detecting the attack strongly correlates with the honeypots conforming to the selection criteria.
\end{enumerate}

Amplifier hit lists may be static, in which case the attack remains invisible if no honeypot is on the list, or dynamic. In the latter case, honeypots may observe scanning and respond accordingly. Honeypots typically expose a low amplification factor due to legal reasons, which makes them less attractive in many attacks.

\paragraph{Finding a good completeness metric}
Often, honeypot platforms have a limited diversity in geography or network topology.  
A valid metric for estimating the completeness of honeypot observations needs to infer global knowledge from local observations, which is the more challenging the smaller and less diverse local observatories are. Preferably, such metric can at least provide a rough estimator of the error inherent to the measurement system. 
As we have seen in the previous discussions, such an indicator cannot be extracted from the pure measurement set alone. Instead, orthogonal sensors and correlating analyses are needed to capture and  quantify the invisible attack data. 

An obvious source of control is to compare with alternate measurements such as flow data, network telescopes, or public attack reports. 
For research that needs to exclusively base on the honeypot platform, we conjecture that additionally observing and analyzing explorative scanning (possibly with varying reply behavior) as well as correlating initial scanning with subsequent attack detection (or not), may open a new angle of view on the completeness of the honeypot attack data.

\section{Completeness}
\label{sec:completeness}

Using additional data sources, we find that honeypots are unable to
observe anywhere near a complete view of real-world attacks, but are
quite good at detecting scanning~activity.

\begin{figure*}
  \begin{subfigure}{.5\textwidth}
    \centering
    \includegraphics[width=\linewidth]{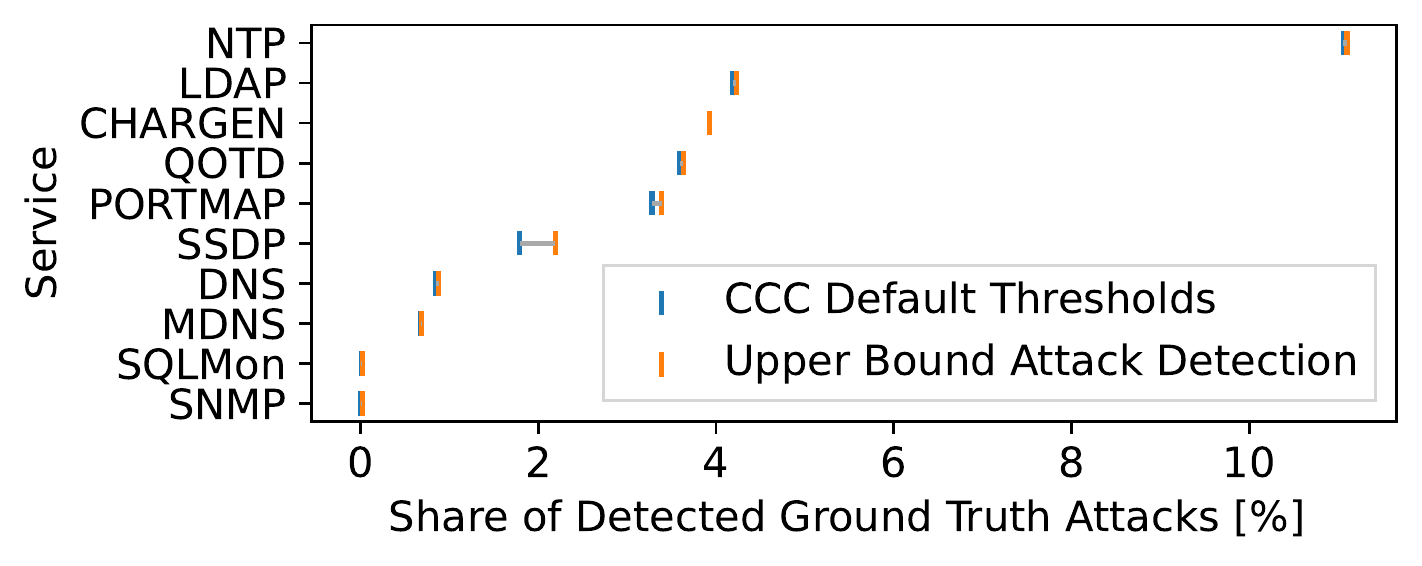}
    \caption{CCC}
    \label{fig:precision:overlap_ground_truth_atks_ccc}
  \end{subfigure}\hfill
  \begin{subfigure}{.5\textwidth}
    \centering
    \includegraphics[width=\linewidth]{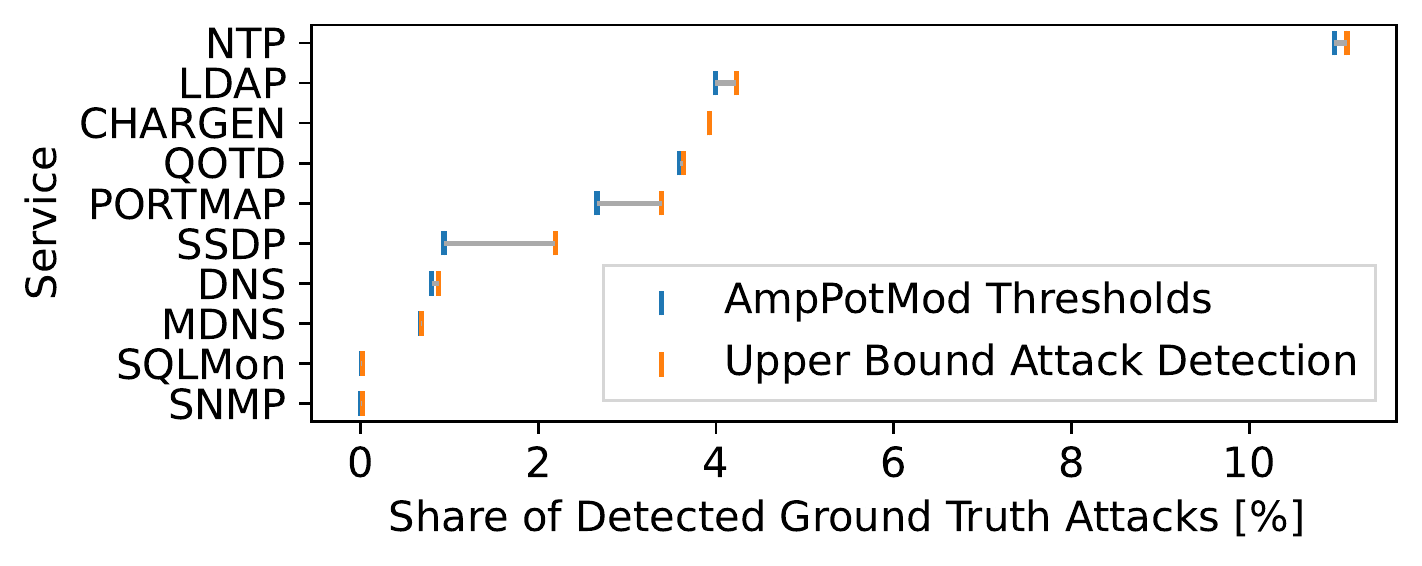}
    \caption{AmpPotMod.}
    \label{fig:precision:overlap_ground_truth_atks_amppotmod}
  \end{subfigure}\hfill
  \caption{Honeypot probes detect at most 11\% of the ground truth attacks.There is no room for fine-tuning the thresholds to improve the detection rate, because the probes simply do not observe more events for the victims.}  
  \label{fig:precision:overlap_ground_truth_atks}
\end{figure*}

\subsection{The Honeypot View is Mostly Incomplete} 
\label{sec:precision:completeness}
Similar to honeypots, our DDoS provider data shows that DNS (60\%), NTP (23\%),
and LDAP (8\%) are the most popular protocols misused for amplification.
Leveraging this
real-world baseline data, we can now independently assess whether the
honeypots grant a reasonably complete view on attacks.  To this
end, we detect attacks using the honeypot data and the default CCC or
AmpPotMod thresholds.  Then, we calculate the share of overlapping
attack events in the DDoS provider baseline data for each protocol.  The
results are visualized in
\autoref{fig:precision:overlap_ground_truth_atks}.  We find very limited
overlap, \ie honeypot views on amplification attacks are mostly
incomplete.  For the best performing protocols, for which we confirmed
uninterrupted operations and convergence in the previous sections, we
only observe 11\% (NTP) and 4\% (LDAP) of attacks.  This is in stark
contrast to current convergence
measurements~\cite{griffioen2021adversarial}, which suggest that we
should observe at least 90\% of NTP attacks with 50 sensors.  Our
results, however, comply with recent findings
(4\%~\cite{nawrocki2021ecosystem} or 8.18\%~\cite{kopp2021ddos}), which
examine the overlap between honeypots and IXPs, but based on baseline
data.
We acknowledge that our baseline data is limited to those networks that
share attack alerts with the DoS provider.  Nevertheless, we want to
stress two important details.  \one our data represents a fairly large
share of the market (up to 22\%) and \two for a complete coverage of all
attacks, honeypots should \emph{at least} observe most if not all of our
baseline attacks.

Notably, the relative popularity of DNS differs between honeypots (see
paragraph above) and other vantage points such as IXP-based measurements
(DNS 43\%, NTP 25\%, LDAP 20\%~\cite{kopp2021ddos}) and our baseline
data (DNS 60\%, NTP 23\% and LDAP 8\%).  We argue that honeypots miss a
substantial portion of DNS attacks for two reasons: \one DNS amplifiers
have the highest churn rates~\cite{kuhrer2014exit, kuhrer2015resolvers},
which makes it necessary for attackers to rescan frequently. Hence,
attackers can easily rotate between amplifiers and prefer new
amplifiers~\cite{nawrocki2021ecosystem}.  \two Although the DNS
ecosystem consists of various amplifiers \cite{nawrocki2021transparent},
the driving factor for amplification are queries for names with large
zones. This means that the attackers can utilize most amplifiers if they
select such a name, which makes the honeypots less attractive or at
least less likely to be used.  This is supported by the fact that DNS
has the slowest convergence \cite{griffioen2021adversarial}.

\subsection{No Potential for Better Attack Thresholds}
\label{sec:precision:better_atk_thresholds}
We ask whether we can fine-tune the thresholds to improve results.
For this, we infer the upper bound of attack detection.  We use the most
permissive thresholds, \ie every event is classified as an attack.  This
potentially leads to many false positives because even scanners sending
just one packet to the honeypot platform will be interpreted as an
attack.

We visualize the results  in
\autoref{fig:precision:overlap_ground_truth_atks}.  The grey horizontal
lines indicate the potentials for improvements.  We find that we cannot
significantly fine-tune the thresholds because the honeypots simply do
not observe any event for the victims in our baseline data, \ie there is
no packet that relates to any of the IP~addresses under attack.

This limited potential suggests that optimizing the thresholds would
lead to overfitting with respect to our baseline data set.  Also, such
thresholds would only be optimal for a particular point in time and
probably lose the acquired precision in the long term.

\subsection{Misclassification of Scans}
\label{sec:precision:scan_thresholds}

We now utilize network telescopes to assess whether attack detection thresholds for honeypots successfully eliminate scan events.

\begin{figure}
  \centering
  \includegraphics[width=\linewidth]{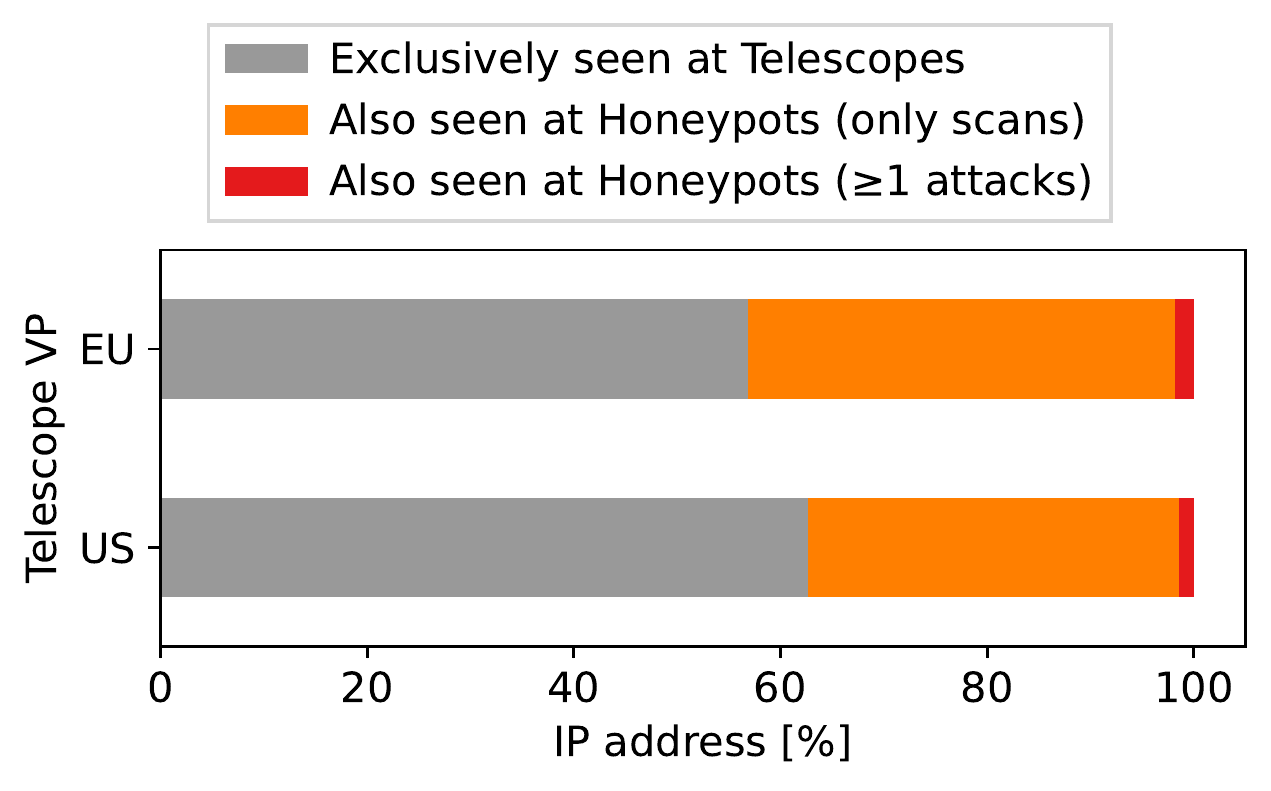}
  \caption{Hosts scanning our telescopes and connecting to our honeypot platform. CCC thresholds mainly infer scanning events, indicating successful scan event detection.}
  \label{fig:overlap_telescopes}
  \vspace{-.55cm}
\end{figure}

\paragraph{Telescopes and honeypots observe the same scanners} At our
telescopes, we identify all scanners contacting service ports supported
by the CCC honeypots.  During our main measurement period, we find 27k
unique scanner addresses in the US and 16k in the EU.  We now check
whether these addresses have been observed at the CCC platform.  The CCC
honeypots observe 37.4\% of the US scanners and 43.1\% of the EU
scanners.  Not all scanners are observed since not every scanner
performs a complete address space scan, \eg it is part of a botnet or
pool which  splits the address space, or operates very
locally~\cite{hiesgen2022spoki}.

\looseness=-1
\paragraph{Telescopes and honeypots agree on scanners}
We now apply the default CCC thresholds, inspect the events caused by
scanners, and compare the results by region in Figure
\ref{fig:overlap_telescopes}.  Strikingly, 36\%  of US scanners
only triggers scan events at the honeypots.  Likewise, the
honeypots infer attacks only for 1.4\% of the scanners.  At the EU, we
observe similar trends with 41.3\% of the sources performing scans only
and 1.8\% triggering attacks.  We repeat this analysis using the
AmpPotMod thresholds and find comparable results (not visualized).
However, AmpPotMod thresholds detect slightly fewer attacks.  The share
of addresses performing attacks decreased for both vantage points, the
US ($1.4\% \rightarrow 1.1\%$) and EU ($1.8\% \rightarrow 1.4\%$).

\looseness=-1
Please note that the detected attacks are not necessarily
misclassifications.  Upon receiving a response, a scanner might start
testing the capabilities of a honeypot, which triggers an attack event.
But such behavior is rather unlikely because scanners try to remain
under the radar in order to avoid being blacklisted and to help
discover as many victims as possible.  Overall, we find that both
threshold configurations are successful in exposing scan events as such.

\section{Network Access, Economic Considerations}
\label{sec:ecosystem}

\subsection{Network Types and Service Proximity}
\label{sec:ecosystem:network}

Honeypots can be deployed in any type of network with public reachability.
Similar to the various threshold configurations, the effect of different
network access types is little understood.  For the large amplification
honeypot platforms, we typically see sensors placed on eyeball, hosting, and
academic networks.  Still, we miss discussion on how the observations
differ across network~types.

Quantitative and qualitative differences have been shown for non-amplifying
honeypots placed in mobile network service providers, darknets, and academic
networks \cite{waehlisch2013mobile}, \eg only a few topological Internet-domains have started to place dedicated focus on attacking mobile networks.
For example, malware and scanners have been shown to limit their operations
geographically and topologically~\cite{hiesgen2022spoki}.  Such 
differences observed across network types must also be anticipated for amplification
honeypots.

Since many open services disappear because of IP churn
\cite{kuhrer2014exit} and not because they were taken down, it is
beneficial to periodically rescan the network to update the service-to-address-mappings. 
According to~\cite{klick2016citizen}, scan traffic can be reduced by
25-90\% while missing only 1-10\% of the population.  This means that attackers
utilizing such optimization will more likely discover and misuse honeypots
that are in proximity to other amplifiers.

Cloud providers share their physical infrastructure through the use of
virtualization.  Outages and the mitigation of (unrelated) attacks) on shared
infrastructure may affect honeypot measurements and thereby attack detection.
Therefore, researchers need to pay close attention to the fate-sharing risks
and factors of an otherwise well-functioning honeypot system.

\subsection{Economic Considerations}
\label{sec:ecosystem:economy}

\looseness=-1
Attackers misusing amplifiers are often operating in pursuit of economic goals.
For instance, fee-based booter (or stresser) operators sell DoS attacks
as-a-service and have been linked to the misuse of open
amplifiers~\cite{santanna2015booters, krupp2017linking}.  Booter operators run
websites where any individual can purchase
attacks~\cite{santanna2016blacklist}.  
Although some DDoS-as-a-service websites have been shown to utilize the same set of amplifiers, for most operating websites the overlap is minimal~\cite{santanna2015booters}.

Researchers have used booter services to attack their own infrastructure and
found attacks utilize on average 346~amplifiers from 27 autonomous
systems~\cite{kopp2019hide}.  Honeypots observed only $\sim$40\%
of DNS self-targeted attacks~\cite{krupp2017linking}.
Overall, booter services are responsible for a significant number of
amplification attacks, \eg 26\% of DNS and 13\%~of NTP attacks were linked to a
specific set of booters~\cite{krupp2017linking}.  This means that observations
by honeypot sensors can be extremely biased if they are used by a specific
booter.  Furthermore, take downs of booter websites can reduce the
number of observed attacks~\cite{collier2019booting, kopp2019hide} and
negatively bias the  attack~landscape perceived by a honeypot system.

Unfortunately, little can be done to influence the selection process of
attackers.  Obviously providing potent amplifiers helps, however, this opposes
ethical measurements which deploy rate limiting.  Therefore, special care has
to be taken while analyzing significant peaks and drops, \eg for number of
attacks for a specific protocol.  Variations in attack detection may rather be caused
by a specific booter omitting the honeypots from active use rather than a
reflection of aggregate attack event trends.

\section{Discussion}
\label{sec:discussion}
\vspace{-0.4cm}

\paragraph{Why our results differ}
Our results on the completeness of honeypot observations clearly differ from past research, indicating that honeypot systems miss a substantial share of all Internet-wide attacks.
We identify two major reasons for the differences:
First, honeypot observations, especially for early deployments, show a very fast convergence, which was misinterpreted as an indicator for completeness. Convergence, however, can only serve as an indicator for cost-efficiency of a particular deployment.
Second, the access to orthogonal vantage points, \eg commercial on-path mitigation appliances, is rare and regulated by NDAs.
By closely cooperating with a DDoS mitigation provider, we designed a method that evaluates the completeness of honeypot observations but still respects data privacy.

Following our systematic approach, we believe that our results exhibit \emph{a more trustworthy} view on the amplification ecosystem.
This is because 
\one we do not select a singular configuration but explore complete threshold spaces and analyze convergence after random permutations of the sensor order, and
\two our completeness results are bolstered  by a curated DDoS attack list from a major mitigation~provider.

\paragraph{Our limitations} Our results are based on a dataset that was
gathered recently and covers a specific time period. Not all
datasets that were used in prior publications were at our disposal; hence we 
could not evaluate some of the  prior research against our baseline data.
Even with the data constraints, however, we were able to use the configurations
of various publications to compare detection properties of multiple honeypot
thresholds.  Furthermore, we argue for our finding that honeypots capture only
a limited part of the global attack landscape since other
work~\cite{krupp2017linking, kopp2021ddos, nawrocki2021ecosystem} has raised
similar concerns, while using complementary vantage points and time periods.

\paragraph{Convergence vs. completeness}
Although convergence does not indicate completeness, both properties can occur at the same time.
In \autoref{tab:conv-vs-compl}, we depict examples under which conditions these properties occur.
Attackers who are able to detect honeypot sensors as their targets, \eg due to rate-limiting at honeypots, can decide to never use them as amplifiers.
This impedes completeness.
At the same time, attackers that repeatedly use the same honeypot sensors for different attacks foster convergence since they add additional weight to the frequency of occurrence.
This illustrates that deploying more honeypots sensors, even with a diverse geographical and topological distribution, does not necessarily lead to more reliable results.
Instead, a thorough understanding of the attacker decision making is essential.

\begin{table}%
\setlength{\tabcolsep}{3.8pt}
  \caption{Convergence does not implicate but can co-occur with completeness, depending on the attacker behavior.}
  \label{tab:conv-vs-compl}
  \centering
   \begin{tabular}{c c p{2.8cm} p{2.9cm}}
    \toprule
      \multirow{2}{*}{\emph{Conv.}} & \multirow{2}{*}{\emph{Compl.}} & 
      \multicolumn{2}{c}{Scenarios} \\
        \cmidrule(lr){3-4} && Honeypot Observations & Reflector Selection \\
    \midrule
        \multirow{2}{*}{\xmark} & \multirow{2}{*}{\xmark} &
            Each sensor captured different attacks. &
            Some attackers did not use the honeypots. \\
    \dhline
        \multirow{2}{*}{\cmark} & \multirow{2}{*}{\xmark} &
            Multiple sensors captured the same. &
            Some attackers did not use the honeypots. \\
    \dhline
        \multirow{2}{*}{\xmark} & \multirow{2}{*}{\cmark} &
            Each sensor captured different attacks. &
            All attackers used the honeypots. \\
    \dhline
        \multirow{2}{*}{\cmark} & \multirow{2}{*}{\cmark} &
            Multiple sensors captured the same. &
            All attackers used the honeypots. \\
    \bottomrule
  \end{tabular}
\vspace{-0.35cm}
\end{table}
\setlength{\tabcolsep}{5.25pt}

\paragraph{Honeypots are useful} It is discouraging to detect only up to 11\% of attacks, in particular when facing the costs of deploying
(renting cloud servers, buying dedicated hardware \etc) and maintaining
amplification honeypots.  Even though honeypots lack a complete view on the
attack landscape, knowing this imperfection removes unwanted interpretation
bias.  We argue in favor of honeypot results as an
important component of a larger complex ecosystem even if they are imperfect.
We believe accepting this will help researchers to better interpret the
observed phenomena and to understand \emph{their fragment} of attacks.  Since
researchers are restricted to ethical measurements and hence rate limit
honeypots, attackers will always be able to elude the trap.

\paragraph{Recommendations for attack definitions} A comprehensible, precise
attack definition is essential for honeypot research.  We find only textual
definitions of attack thresholds in related work.  Although these can be
sufficient, they are ambiguous and open to interpretation.  This is why we
recommend precise wording, preferably taken from common sources.  A good
candidate for this is found in the IPFIX specification \cite{RFC-5102}, from
which we adopted the usage of \emph{idle timeout} and \emph{flow}.

Overall, at least three definitions are required: \one What identifies a
victim? This could be a single source IP address, an IP~prefix, an autonomous
system, a name \etc \two What is an attack flow? One should clarify which flow
keys are observed for flow inference and which thresholds are applied.  \three
What is an attack (event)? This is especially important for system-wide flow
identifiers, when distinct attack flows towards the same victim are observed
from different vantage points.

\paragraph{Directions for the future}
Our results indicate the importance of  extensive baseline data and 
ground truth.  However, our community should not depend on it.
Non-proprietary, auxiliary vantage points such as telescopes and correlating observations can also help to
assess or improve the precision of measurements.  We see such heterogeneous
deployments in active use by commercial parties, \eg GreyNoise.  Simply
adding more honeypot sensors does not necessarily solve measurement challenges
such as the honeypot convergence, which is among other potential obstacles
caused by the decision making process of the~attacker.

A fundamental problem for honeypot research is that aggressive scans exhibit
traffic patterns similar to reflective amplification attacks.  Conversely,
low-volume attacks misusing relatively few amplifiers can resemble 
patterns of cautious scanners.  Prior work was conducted based on the assumption that these
phenomena do not intersect, but they do.  This intersection can be illuminated
by considering a careful definition and explanation of thresholds w.r.t. the observed data and the current amplification ecosystem.

The ever-changing ecosystem is the reason why we refuse to recommend a single
best threshold configuration in this paper.  It is likely that any such
recommendation will soon be obsolete as attacks and methods evolve.
Additionally, our results indicate that even perfect attack classifications
will not be able to detect all attacks.
There is room for clarification on the impact of thresholds, and the
correlation of minor events to make classification of various measurements
easier.
However, opportunistic classifications into obvious scans and obvious attacks
are valuable.

With these considerations in mind, we go beyond just a call for comparable
metrics.  Given the same dataset, we need a way to compare the effects of
different thresholds.  We also encourage authors to present
detailed analysis on their choice of attack thresholds.
Finally, since the observation range of honeypots is directly related to being targeted by attackers, we argue that
a future research agenda should include methods to replicate the creation of amplifier hit lists.
Mimicking this part would complement our tool set and improve informed honeypot deployment.

\section{Conclusion and Outlook}
\label{sec:conclusion}
\vspace{-0.3cm}

In this paper, we revisited methods to measure and infer reflective
amplification attacks based on honeypots.  We applied a data-driven approach
that allowed us to challenge long-held assumptions.
Using data from a large-scale honeypot, multiple network telescopes, and
extensive baseline data from a leading DDoS mitigation provider, we were able
to reproduce, confirm, or disprove common measures of attack detection,
honeypot convergence, and attack completeness.

Contrary to popular belief, we found that \one honeypot convergence has limited
significance because it is a statistically unstable metric and \two
observations by honeypots are incomplete, honeypots miss large fractions of
ground truth attacks.  We explored the complete spectrum of attack detection
thresholds and embedded the thresholds of related work in our system.  Related
work, although using different thresholds, largely produces comparable results
but common thresholds cover only a very narrow part of the parameter space.  We
highlighted the various features that should be considered by researchers when
deploying honeypots and analyzing data.  These include setup properties, flow
identifiers, and attack thresholds.

Our results underscore three open challenges. %
First, a well-defined definition of an attack, which accounts for
traffic patterns observed by honeypots.  Second, a reliable metric to assess
the completeness of honeypot observations.  Such metric should provide an error
margin and not depend on external ground truth data.  Third, to increase
completeness, well-defined features that guide honeypot deployment.  These
might include deployments in heterogeneous network types, better protocol
emulation, or sophisticated rate limiting methods.
Most importantly, our community should gain a better understanding of the mechanics behind the creation of amplifier hit lists.
Being able to reproduce the set of amplifiers used by attackers will allow researchers to tailor amplification honeypots in terms of location and behavior such that they will be targeted and capture a sufficiently complete view.

\paragraph{Acknowledgment}
We would like to thank our shepherd Jianjun Chen and the anonymous reviewers for their very detailed and valuable feedback, which helped to improve this paper.
We gratefully acknowledge the DDoS mitigation provider and the CCC honeypot team for providing data.
This work was partly supported by the German Federal Ministry of Education and Research (BMBF) within the project PRIMEnet.

\label{lastbodypage}

\balance

\bibliographystyle{IEEEtran}
\bibliography{bibliography}

\appendices
\section{Examining Convergence for LDAP}
\label{apx:ldap_convergence}

In \autoref{fig:precision:convergence_ldap} and \autoref{fig:precision:ldap}, we present detailed results of the convergence measure for LDAP.
These results confirm that our convergence observations for NTP presented in \autoref{sec:convergence:reproducing} and \autoref{sec:convergence:fair_introspection} also held for LDAP.

\textblockcolour{white}
\setlength{\TPboxrulesize}{0.0pt}
\begin{textblock*}{\textwidth}(2cm,17cm)

\begin{figure*}
  \begin{subfigure}{.5\textwidth}
    \centering
    \includegraphics[width=\linewidth]{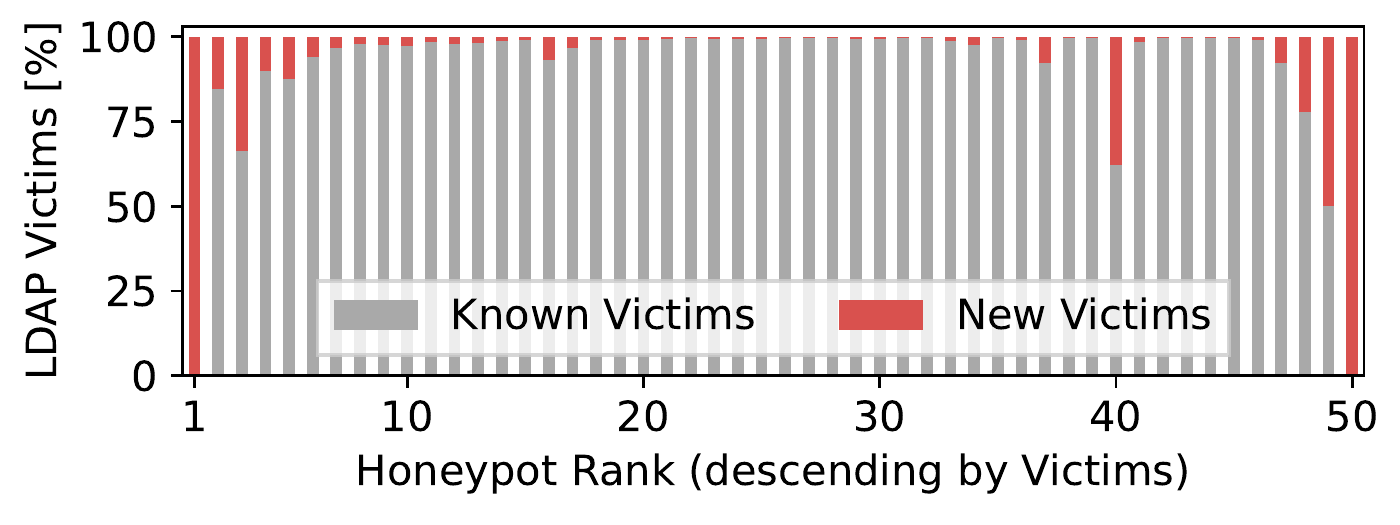}
    \caption{New victims per sensor.}
    \label{fig:precision:convergence_ldap_sbars}
  \end{subfigure}\hfill
  \begin{subfigure}{.5\textwidth}
    \centering
    \includegraphics[width=\linewidth]{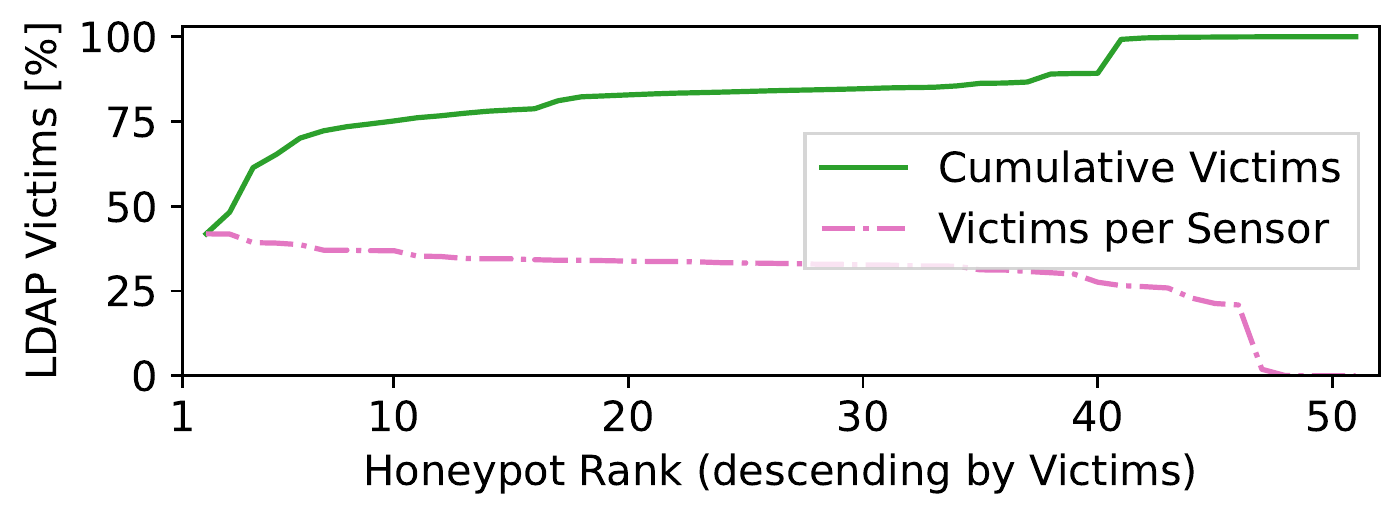}
    \caption{Victims per sensor and cumulative total.}
    \label{fig:precision:convergence_ldap_cumsum}
  \end{subfigure}\hfill
  \caption{Convergence behavior for LDAP using a near-optimal selection of honyepot sensors.}
  \label{fig:precision:convergence_ldap}
\end{figure*}

\bigskip

\begin{figure*}
  \begin{subfigure}{.5\textwidth}
    \centering
    \includegraphics[width=\linewidth]{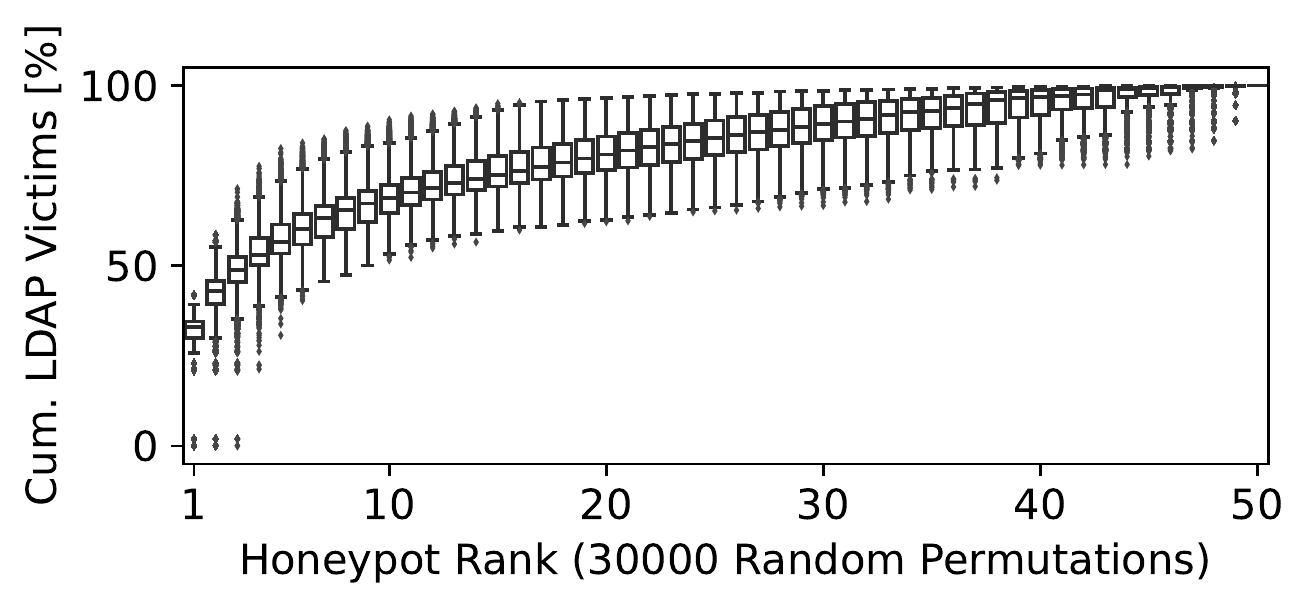}
    \caption{Convergences. High variances in the results suggest that \\ convergence is less stable than previously assumed.}
    \label{fig:precision:permutations_ldap}
  \end{subfigure}\hfill
    \begin{subfigure}{.5\textwidth}
    \centering
    \includegraphics[width=\linewidth]{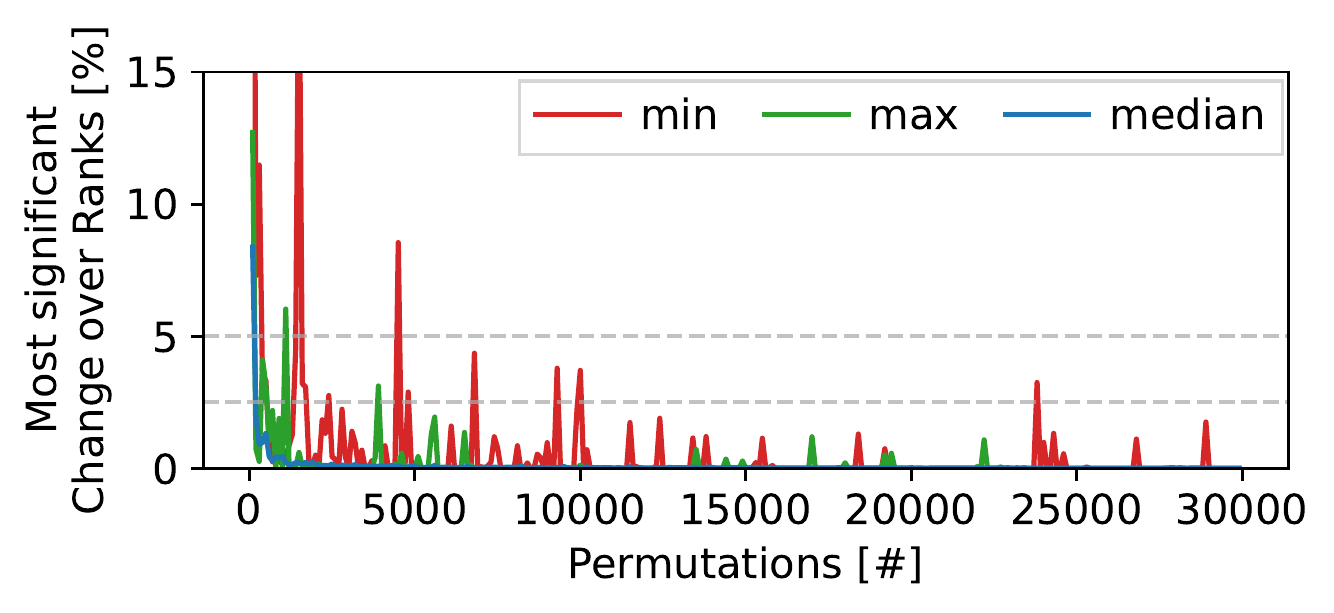}
    \caption{Relative differences of min, med, and max of detected victims. Even at $\approx$25k, worst-case results (min) differ by less than 2\%.}
    \label{fig:precision:nperm_diff_ldap}
  \end{subfigure}\hfill
  \caption{Examining the convergence of LDAP over 30k permutations.}
  \label{fig:precision:ldap}
\end{figure*}

\end{textblock*}

\label{lastpage}
\end{document}